\documentclass[prb,twocolumn]{revtex4}
\usepackage{graphicx}
\usepackage{dcolumn}
\usepackage{bm}
\usepackage[english]{babel}
\usepackage{color}

\begin{document}
\pdfoutput=1

\preprint{}

\title{Exciton-photon interactions in semiconductor nanocrystals: {radiative transitions, non-radiative processes,} and environment effects}

\author{Vladimir A. Burdov}
\affiliation{N. I. Lobachevsky State University of Nizhny Novgorod, 23 Gagarin avenue, 603950 Nizhny Novgorod, Russian Federation}

\author{Mikhail I. Vasilevskiy}
\affiliation{Centro de F\'{\i}sica, Universidade do Minho, Campus de Gualtar, Braga
4710-057, Portugal}
\affiliation{International Iberian Nanotechnology Laboratory, Braga 4715-330, Portugal}

\date{\today}

\begin{abstract}

In this review we discuss several fundamental processes taking place in semiconductor nanocrystals (quantum dots, QDs) when their electron subsystem interacts with electromagnetic (EM) radiation. The physical phenomena of light emission and EM energy transfer from a QD exciton to other electronic systems such as neighbouring nanocrystals and polarisable 3D (semi-infinite dielectric or metal) and 2D (graphene) materials are considered. The cases of direct (II-VI) and indirect (silicon) band gap semiconductors are compared. We also cover the relevant non-radiative mechanisms such as the Auger process, electron capture on dangling bonds and interaction with phonons. The emphasis is on explaining the underlying physics and illustrating it with calculated and experimental results in a comprehensive, tutorial manner.     

\end{abstract}


\maketitle
\section{Introduction}

Semiconductor nanostructures form a basis for modern electronic technologies. Nowadays,~they are employed in a wide range of applications in the fields of optoelectronics, photonics, photovoltaics, biosensing, photocatalysis, {etc}. By virtue of the quantum confinement effect, their electronic spectra and, consequently, the related optical properties depend on the nanostructure size. This feature is most pronounced for zero-dimensional (0D) objects, nanocrystals (NCs), where the conduction band electron motion is fully localized in all directions. Consequently, at least a part of the NC energy spectrum is completely discrete. In the limiting case of strong quantum confinement effect, when the NC size is much less than the effective exciton Bohr radius, the former strongly influences the electron and hole energies. It leads to the size-dependent energies of photons emitted or absorbed by the nanocrystals and allows one to control their optical spectra. For this reason, they sometimes are called ``artificial atoms''~\cite{Ashoori96}. A more scientific term to distinguish semiconductor NCs with strongly size-dependent electronic and optical properties is Quantum Dot (QD), which will be used in this article.\footnote{Let us note that the term Quantum Dot was originally proposed for a lithographic lateral nanostructure based on an AlGaAs/GaAs heterostructure
  with 2D electron gas, probably by M. A. Reed in 1986. Beyond such structures
  and isolated nanocrystals, it covers also self-assembled epitaxial QDs and
  small transistor-like structures where discrete energy levels are created by
  electrostatic confinement. Here we shall employ this term just for
  semiconductor nanocrystals.} 

After the pioneering works by Efros and Efros~\cite{Efros-Efros} and Brus~\cite{Brus83} explaining the origin of size-dependent optical spectra of nanocrystals, the interest to these objects grew exponentially through the 1980s and 1990s.
A variety of methods has been used for preparation of crystallites with sizes not greater than several nanometers, such as ion implantation~\cite{Kanemitsu,Zhuravlev,NIMB}, chemical vapour deposition from a gas phase,~\cite{Negro} magnetron sputtering~\cite{Tsybeskov}, colloidal synthesis~\cite{Nayfeh}, electron beam epitaxy~\cite{Meldrum}, {etc}. One of these methods, chemical growth in colloidal solutions, allows for obtaining good quality NCs of a broad range of semiconductors with an almost spherical shape and a rather narrow size distribution, characterized  by rather narrow emission bands, which is, as a rule, extremely desirable for applications. This method was first suggested for II--VI NCs by Murray {et al.}~\cite{Murray} and became widely employed later (see, e.g., chapters by Kudera {et al.}, Reiss, Gaponik and Rogach in the book~in \cite{Rogach2008} for review). The colloidal chemistry methods work fairly well for II--VI and III--V materials such as CdSe, CdS, CdTe, InAs, InP, {etc.} {The authors of}~\cite{Rogach2008,Bruchez98} and several other groups have been able to produce good QDs of some IV--VI materials~\cite{Voznyy2017,PbSe-QD} and silicon QDs (see,~e.g., reviews in ~\cite{KDoh,McVey}).  

Colloidal QDs of II--VI materials (especially CdSe and CdTe) are probably the most  studied. During the last decade, a considerable research activity has been focused on the  synthesis and investigation of the optical properties of ZnSe QDs. These nanocrystals exhibit large blue shift of the photoluminescence~\cite{Memon} and high enough (up to 50~percent) quantum yield~\cite{Senthil}. Similarly to CdSe NCs, the ZnSe QDs can be synthesized with a sufficiently narrow size distribution, which is possible to control by temperature~\cite{Baum}. Due~to ZnSe's lower toxicity, compared to many other II--VI semiconductors, these QDs turn out to be attractive objects also for biosensing~\cite{Moura}.

Further improvement of the light emission properties of such QDs was achieved by fabrication of core--shell structures, successful for several pairs of II--VI~\cite{Talapin2001} and IV--VI~\cite{SMaiti} materials, where the shell made of a wider band gap material provides a better protection of the quantized electronic states in the QD core~\cite{Rogach2008}. For ZnSe/ZnS core--shell QDs it was revealed, in particular, that thermodynamical (slow) growth of the ZnS-shell on the colloidal ZnSe-core leads to the quantum yield increase because of decreasing amount of the traps at the core--shell interface~\cite{Ji}.    

The maturity of this technology of synthesis of semiconductor nanocrystals\footnote{We are not aware of successful colloidal chemistry synthesis of matrix-free
  silicon nanocrystals. However, they can be fabricated by other
  methods \cite{Pavesi,Pereira2013,Si-NC2020} and processed in a way similar to
  colloidal NCs of other semiconductors. \cite{Holman2010} They will be
  discussed below.} is witnessed by the incorporation of colloidal QDs into real-world products such as colour displays~\cite{QD-TV} and light-emitting diodes for lighting~\cite{QD-LED}, both already commercialized by large companies such as Samsung~\cite{Samsung} and OSRAM~\cite{OSRAM}, respectively (see Ref.~\cite{Zhaojun2020} for a recent review covering both colloidal and epitaxial QDs).
More on the scientific research side, a broad variety of nanostructures can be prepared using colloidal NCs as building blocks, in particular, multilayer structures of QDs of different average size, deposited on different substrates~\cite{Rogach2008}. Combining these structures with other materials, such as organic dielectrics~\cite {Basko2000}, epitaxial quantum well (QW) heterostructures~\cite{Achermann2004}, metallic nanoparticles~\cite{Knorr2013,Lunz2012}, patterned metallic surface~\cite{Pompa2006} or graphene~\cite{Chen2010,Agarwal2013} can result in new interesting effects and applications, which physics is related to the coupling of QD excitons with elementary excitations (such as surface plasmons or QW excitons) in the surrounding materials. One interesting possibility is the excitation of colloidal QDs by pumping energy through a nearby epitaxial QW, as demonstrated in~\cite{Achermann2004}.
Another research topic with QDs, popular in the past two decades is the modification of their emission properties caused by the strong light--matter interaction, which has been achieved for a variety of composite structures that include point emitters embedded in Fabry--Perot microcavities, micropillars and photonic crystals as recently reviewed in \cite{Dovzhenko2018}.

In this article, we overview several aspects of the exciton--photon interactions in nanocrystal QDs with focus on the role of non-radiative processes and environment, which affect the light emission. The QD photoluminescnce is a result of competition between the radiative and various non-radiative processes, such as capture on dangling bonds, multi-phonon intra-band relaxation, Auger recombination, {etc}.
In the past decade, a growing interest has been paid to the multi-exciton dynamics in nanocrystals because of their high potential for photovoltaic applications. Such processes as carrier multiplication (or multi-exciton generation) in nanocrystals, as well as the Auger recombination (which is just a fast reverse process), are widely discussed in the literature~\cite{SMaiti,Klimov,Bruhn,Haverkort} and we will consider these effects below. Special attention is traditionally paid to nanostructured silicon~\cite{Pavesi,Khriach,Ray,Barba,Priolo1} because of its widest use in microelectronics, high purity, natural abundance, low cost, and non-toxicity; we shall also dedicate some space to the exciton--photon interactions and main non-radiative processes mentioned above, in Si NCs and related nanostructures.

As a rule, one deals with nanocrystal {\it {ensembles}} (rather than with isolated QDs). Therefore, non-radiative energy exchange between the nanocrystals is possible {via} the F\"{o}rster-type~\cite{Forster} and Dexter-type~\cite{Dexter1953} exciton migration. Both mechanisms were originally proposed for fluorophores. The second one, based on the exchange interaction between electrons located on different sites can be neglected in the presence of allowed dipole transitions~\cite{Dexter1953}.
Recently, a third, so-called exciton tandem tunneling mechanism was proposed~\cite{Reich}, which is specific of connected nanocrystals. 
The most universal process named F\"orster Resonant Energy Transfer (FRET) was first observed for QDs by Kagan~{et~al.} in specially designed films containing two different sizes of nanocrystals acting as donors and acceptors, respectively~\cite{Kagan96}. 
Later it was shown in a number of works~\cite{Crooker2002,Franzl2004,Lunz2010,Lunz2011,Yu,Kawazoe} performed on systems composed of two different QD species that the luminescence of the smaller dots (donors) is quenched by the large dots (acceptors), whose emission in turn is enhanced. These studies demonstrated a dependence of the FRET effect on the NC density and spatial arrangement.
Results of other experiments performed on multilayer SiO$_x$/SiO$_2$ structures~\cite{Linnros,Heitmann,Glover}, porous Si~\cite{Ben}, three-dimensional (3D) ensembles of Si~\cite{Priolo,Balberg} crystallites were interpreted as a  manifestation of exciton migration {via} FRET-type mechanisms.
The~effect has a potential interest for photonics (e.g., photoluminescence upconversion in QD ensembles~\cite {Santos2008}), sensing~\cite {Nabiev2004}, lighting and energy harvesting (e.g., by unidirectional energy transfer in size-gradient layered QD assemblies~\cite {Crooker2002,Franzl2004} or fractal aggregates~\cite {Sukhanova2006,Bernardo2014}). We shall discuss it below, in particular, addressing the question whether the FRET rates and length scales can be tuned by the photonic environment~\cite {Rabouw2014}.

The influence of the environment on the exciton-related optical properties of NCs is the third topic that will be discussed in this overview. The most common effect is the energy transfer between a photoexcited QD and surrounding materials, which can be reversible or not. It has the same physical nature as FRET. In most cases, it is responsible for the photoluminescence (PL) quenching \cite{Chance,Koppens2011}.
However, the environment can be used for engineering the photonic density of states (DOS) in the vicinity of the QDs, often referred to as the Purcell effect~\cite{Purcell}. Observed for the first time in QDs about twenty years ago~\cite{Gerard_PRL1998}, it has recently been discussed with respect to radiative decay rates of embedded point emitters~\cite{Senden2015}. Moreover, strong near-field effects associated with localized surface plasmon resonance (LSPR) in metallic (nano-)structures can enhance PL emission~\cite{Pompa2006,Shimizu2002}.   
Moreover, one can think of energy transfer from a recombining QD exciton to propagating surface plasmons that would carry the energy over a large distance and then eventually transfer it to another QD (by creating an exciton), thus allowing for a long-range exciton transport between two dots, much more efficient than if it occurred directly. Below we consider the influence of a flat interface between two media on the PL emission and FRET rates for a QD emitter located in the vicinity of the interface~\footnote{We will not cover here the broad topic of interactions between QDs and metallic nanostructures supporting {\it localized} surface plasmons because it deserves a separate article. Refs.~\cite{Lunz2012,Knorr2013,Schreiber2014,Rakovich2015} can provide an introduction to this topic.}.

The article is organized as follows. In Section \ref {sec_RT}, we introduce basic notions of radiative transitions and discuss their rates (i.e., probabilities per unit time) for NCs of materials possessing direct or indirect band structure. {In particular, the rates are calculated for intrinsic and doped silicon NCs, which are either hydrogen-coated or halogenated.}~Section~\ref {sec_NRT} is devoted to non-radiative transitions and considers the Auger recombination, dangling-bond traps and phonon-assisted relaxation of hot carriers. In Section \ref {sec_MEG}, multiple exciton generation initiated by a highly excited electron or hole is considered. Section \ref {sec_FRET} is dedicated to the exciton transfer processes between two QDs {via} the FRET mechanism.
In Section \ref {sec_env}, emission decay and FRET rates near a plane interface between two dielectrics or a dielectric and a metal are discussed. The last section is left for summary and conclusions.    

\section{Light Emission in Nanocrystals}
\label {sec_RT}

\subsection{Spontaneous Emission Rate}
The emission rate of a point dipole emitter located in an infinite dielectric medium with the permittivity $\varepsilon _1$ is given by \cite{Hecht-Novotny} 
\begin{equation}
\gamma _0 =\frac {4k_1^3\vert \mathbf d \vert ^2}{3\varepsilon _1 \hbar}\, ,
\label {eq_gamma0}
\end{equation}
 where $k_1=\sqrt{\varepsilon_1}\omega/c$ is the modulus of the wavevector, $\omega$ is the oscillation frequency, $c$ is the velocity of light in vacuum, $\hbar$ denotes the Planck constant and {$\mathbf d$} is the transition dipole moment matrix element. This expression, with some adaptation, describes the radiative decay rate of an exciton in a QD made of a direct band gap semiconductor. We shall write it as the inverse of the radiative lifetime: 
 
\begin{equation}
\tau_{R}^{-1} \equiv \gamma _0^{(QD)} = \frac{4e^2 E_{if}|{\bf p}_{if}|^2\kappa\sqrt{\varepsilon_1}}{3m_0^2\hbar^2c^3}\, .
\label {eq_tauR}
\end{equation}

Here, $m_0$ and $-e$ are the free electron mass and charge, respectively, and $E_{if}$ and $\mathbf p_{if}$ denote the transition energy between the initial and final states and the momentum matrix element, respectively. The latter is more natural to use in the electronic band theory of crystals; it can be related to $\mathbf d$ appearing in Equation (\ref {eq_gamma0}) as $\mathbf p =-i m_0\omega \mathbf d/e$.
The parameter $\kappa$ in Equation (\ref {eq_tauR}) arises due to the difference in the dielectric constants of the crystallite ($\varepsilon$) and its surrounding ($\varepsilon_{1}$) and takes account of the so called depolarization effects \cite{Jackson}. Hereafter, we consider the crystallite as a sphere with the radius $R$ and dielectric constant $\varepsilon$ embedded in a homogeneous medium with the dielectric constant $\varepsilon_{1}$. In this case, $\kappa$ is given as~\cite{Thranhardt}
\begin{equation}
\kappa = \frac{9\varepsilon_{1}^2}{(2\varepsilon_{1}+\varepsilon)^2}.
\label {kappa}
\end{equation}

This parameter varies within a wide range of values depending on the dielectric constants of the two materials.

\subsection{Nanocrystals of Direct-Band-Gap Materials}

The majority of III--V and II--VI semiconductors  are direct band gap materials. In this case, in the framework of the envelope function approximation,
\begin{equation}
{\bf p}_{if} = {\bf p}_{cv}\langle F_{c,i}|F_{v,f}\rangle,
\label {eq_Pif}
\end{equation}
where ${\bf p}_{cv}$ is the standard bulk-like momentum matrix element of inter-band transitions, and $F_{c,i}$ and $F_{v,f}$, in case of the photon emission, stand for the envelope functions of the initial state in the conduction band and the final state in the valence band, respectively. The~scalar product of these envelope functions determines selection rules for such transitions.

In the simplest case, the QD is treated as an infinitely deep spherical potential well with radius $R$ and the conduction and valence bands are assumed simple and parabolic. Consequently, the electron and hole ground states are described by the same envelope function and the scalar product in (\ref {eq_Pif}) equals to unity. It means that the radiative recombination rate turns out to be almost independent of the QD radius. This dependence manifests only in the transition energy, $E_{if}$. In the limiting case of a strong quantum confinement regime, where $R\ll a_{ex}$ ($a_{ex}$ denotes the effective exciton Bohr radius), the exciton binding energy can be neglected. Then, the model of infinitely deep spherical potential well yields $E_{if}(R) = E_g^{\infty} + \hbar^2\pi^2/2\mu R^2$, where $E_g^{\infty}$ is the band gap energy of the bulk semiconductor and $\mu$ is the reduced effective mass of the electron--hole pair. 

The above assumption concerning the extreme strong confinement regime is not quite realistic; usually the QD radius can be just slightly smaller than $a_{ex}$ (for instance, $a_{ex}$ = 5.6~nm for CdSe~\cite{Rogach2008}) and the exciton binding energy can be estimated as~\cite{Brus83} $E_b\approx 1.79e^2/(\varepsilon R)$. Furthermore, the most common semiconductors have a degenerate valence band with light and heavy hole sub-bands, so that the confined holes are described by two envelope functions, which are completely different from the electron's one~\cite {Efros96}. Thus, the matrix element (\ref {eq_Pif}) does depend on $R$ (see, e.g., chapter by Vasilevskiy in the book~\cite {Rogach2008}). Still, the dependence of $\tau_{R}^{-1}$ on the QD size is rather weak for direct-gap materials as compared to, e.g., silicon NCs considered below. Using typical parameters for CdSe QDs, $\varepsilon \approx 10$, $\varepsilon_1 \approx 2$ and ${\bf p}_{cv}^2/m_0 \approx 10$ eV~\cite{Efros96},
\begin{equation}
\tau_{R}^{-1}[\text{s}^{-1}]\approx 3\times 10^8 E_{if}(R)\, .
\end{equation}

Thus, $\tau_{R}$ is of the order of a nanosecond, in accordance with experimental observations.

\subsection{Silicon Nanocrystals}

{Since the 1990s, the optical properties of Si NCs have been widely discussed~\cite{Pavesi,Khriach,Ray,Barba,Priolo1,Holman2010,FTT,Pereira2013,Si-NC2020}{.} Many methods have been proposed to increase their emissivity, such as doping with shallow impurities~\cite{Tetel,Fujii,FTT1,Surface,JPCM,Nomoto,Klimesova}{,} growth in different matrices~\cite{Kim,Klangsin}}{,} plasma~\mbox{\cite{Nozaki,Zhou}}{,} and colloidal solutions~\cite{Baldwin,Wolf,Bell,Zhou1,Carroll}.~As a result, numerous theoretical predictions and experimental observations of enhanced PL intensity~\cite{Tetel,Fujii,JPCM,Nomoto}, radiative recombination rates~\cite{Klimesova,PRB,Kusova,Ma,Ma1,Wang,Light,Poddubny,JPCC,JAP1} and quantum efficiency of photon generation~\cite{Sangg} have been reported.

It is well known that introduction of shallow impurities is capable of modifying electronic properties of bulk silicon. Similarly, doping with shallow impurities influences the electronic structure of silicon NCs~\cite{PLA,NRL,JPCM08,Chelik,Oliva,Arduca,Nomoto1,IMarri,Hiller}, which, in turn, affects the electron--hole radiative recombination. It was revealed that doping of Si nanocrystals with P or Li is (under certain conditions) capable of improving their emittance~\cite{Fujii,JPCM,Klimesova,Yang}. In particular, the performed calculations~\cite{JPCM,PRB,JPCC,JAP1} show that doping with P or Li can essentially increase the radiative transition rates.

For NCs with $R\gtrsim 1$ nm, semi-empirical methods are usually employed for computing their electronic structure, such as envelope function approximation~\cite{Hybertsen,Yassi,AOT} or tight binding model~\cite{Delerue}. The performed calculations of the phonon-assisted radiative recombination rates in Si crystallites in the SiO$_2$ matrix yielded the values varying from $\sim 10^2$ to $\sim 10^5$ $\text s ^{-1}$ as the crystallite radius decreases from 3 to 1 nm. Within a simple model of an infinitely deep spherical potential well, the dependence of the radiative decay rate on the crystallite radius is  $\tau_{R}^{-1} \propto R^{-3}$~[\onlinecite{AOT}]. For the no-phonon radiative transitions, the rates turn out to be much lower, by one to three orders of magnitude with increasing the size within the same range. In this case, $\tau_{R}^{-1}\propto R^{-8}$~[\onlinecite{AOT}],  i.e., the rate sharply drops as $R$ increases. At the same time, doping with phosphorus provides a several orders of magnitude increase of the radiative recombination rates for the no-phonon transitions. The rates become even 1--3 orders greater than those of the phonon-assisted transitions. They slowly decrease with increasing the NC size, especially for high concentration of phosphorus~\cite{PRB}.

\begin{figure}[t]
  \includegraphics[scale=0.9]{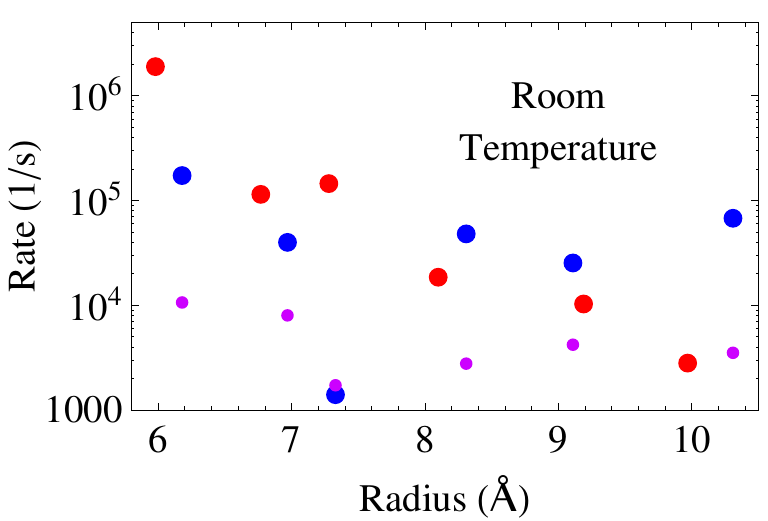}
  \caption{Calculated rates of radiative transitions for undoped (small violet dots), P-doped (big blue dots) and Li-doped (big red dots) Si nanocrystals.} 
  \label{figure1}
\end{figure}

As calculations show, the radiative transitions including both no-phonon and phonon-assisted ones become faster in small Si crystallites. However, for crystallites with $R\lesssim 1$ nm, all semi-empirical methods of calculation turn out to be essentially less accurate. In this case, first-principle calculations have to be carried out. In Figure \ref{figure1}, {the radiative recombination rates computed at room temperature $T$ within the Casida's version~\cite{Casida} of the time-dependent density functional theory (TDDFT) are depicted for H-passivated P-doped (Si$_{46}$H$_{60}$P, Si$_{70}$H$_{84}$P, Si$_{86}$H$_{76}$P, Si$_{122}$H$_{100}$P, Si$_{166}$H$_{124}$P, Si$_{238}$H$_{196}$P), Li-doped (Si$_{42}$H$_{64}$Li,} Si$_{66}$H$_{64}$Li, Si$_{82}$H$_{72}$Li, Si$_{106}$H$_{120}$Li, Si$_{172}$H$_{120}$Li, Si$_{220}$H$_{144}$Li), and undoped (Si$_{47}$H$_{60}$, Si$_{71}$H$_{84}$, Si$_{87}$H$_{76}$, Si$_{123}$H$_{100}$, Si$_{167}$H$_{124}$, Si$_{239}$H$_{196}$) Si nanocrystals.~Here, NCs are considered in vacuum, therefore, $\varepsilon_{1}=1$, while for Si $\varepsilon$ = 12. In this case, the rate values become $\sim$16 times lower compared to the case of Si NCs embedded in silicon dioxide due to the factor $\kappa\sqrt{\varepsilon_{1}}$.

As seen in Figure \ref{figure1}, generally, doped crystallites demonstrate 1--2 orders of magnitude faster transitions than intrinsic  ones. For Si crystallites doped with phosphorus, increasing rates of the radiative transitions have been explained by the efficient mixing of the electronic states in the $\Gamma $ and X points of the Brillouin zone ($\Gamma$-X mixing) caused by the short-range field of the phosphorus ion~\cite{JPCM,PRB}. Meanwhile, the radiative decay in the Li-doped crystallites becomes faster due to the high density of states above the inter-band energy gap (in the range $\hbar\omega - E_g\lesssim k_B T$, where $k_B$ is the Boltzmann constant)~\cite{JAP1}.

Surface chemistry is another method capable of modifying electronic structure of nanocrystals~\cite{QLi,Heuer}. This method is especially efficient for small crystallites, where the role of surface rises. It has been demonstrated experimentally~\cite{Light,Wheeler,Purkait} that chemical synthesis allows to fabricate Si nanocrystals with various ligand coatings, which emit light in visible range with typical nanosecond radiative lifetimes~\cite{Light,De,Sineln}. Calculations based on the tight-binding model~\cite{Poddubny} performed for  CH$_3$-capped Si nanocrystals confirmed the conclusion on the increase of the radiative recombination rate due to the formation of direct-like electronic structure of the crystallite.

Let us now consider some other kind of the surface reconstruction of the crystallites, the halogenation. Table 1 presents calculated rates of radiative transitions between the highest occupied molecular orbital (HOMO) and the lowest unoccupied molecular orbital (LUMO) in completely halogenated with {Cl-, Br- and H-coated} silicon crystallites.

\vspace{0.25 cm}
\noindent {\bf Table 1.} Calculated rates (s$^{-1}$) of the radiative HOMO-LUMO transitions in {H-, Cl-, and Br-passivated} silicon nanocrystals.

\begin{tabular}[c]{cccc}
  \hline
   & X = H & X = Cl & X = Br \\ \hline
  Si$_{35}$X$_{36}$ & $3.7\times 10^5$ & 3.3 & 2.7 \\
  Si$_{59}$X$_{60}$ & $2.9\times 10^4$ & $4.2\times 10^4$ & $3.0\times 10^2$ \\
  Si$_{87}$X$_{76}$ & $1.6\times 10^3$ & $2.4\times 10^2$ & 36 \\
  Si$_{123}$X$_{100}$ & $3.6\times 10^3$ & $4.2\times 10^3$ & $1.7\times 10^5$ \\
  Si$_{147}$X$_{100}$ & $2.2\times 10^4$ & $4.3\times 10^2$ & 15 \\
  Si$_{163}$X$_{116}$ & $6.1\times 10^3$ & $3.0\times 10^4$ & $3.1\times 10^3$ \\
  Si$_{175}$X$_{116}$ & $3.5\times 10^4$ & $1.3\times 10^2$ & $3.5\times 10^2$ \\
  Si$_{317}$X$_{172}$ & $1.4\times 10^4$ & $2.4\times 10^3$ & 63 \\ \hline
\end{tabular}

\vspace{0.25 cm}

It is possible to see from Table 1 that the surface halogenation significantly slows down the radiative recombination in silicon nanocrystals, especially in the case of the bromine coating. Such a slowing down is due to the spatial separation of the electron densities of the HOMO and LUMO states, which takes place in halogenated Si crystallites~\cite{JLett,PCCP,JETP}. The~separation turns out to be more pronounced in the Br-passivated crystallites, in which the HOMO states are strongly concentrated near the Br atoms. Presumably, the bromine atoms produce strong lattice distortions, and the stress fields have an additional localizing effect on the electron density. An effect of the rate reduction was also found for completely fluorine-passivated Si nanocrystals~\cite{Ma1}.

\section{Nonradiative Processes}
\label {sec_NRT}

As is well known, the emissivity of any light-emitting entity depends not only on the radiative transitions' rate, but also on the speed of nonradiative decay processes. In the case of NCs, these include Auger recombination, capture of photo-electrons on dangling bonds and cooling of hot carriers. Below, we briefly discuss all these processes.

\subsection{Auger Recombination}

\begin{figure}[t]
  \includegraphics[scale=1.1]{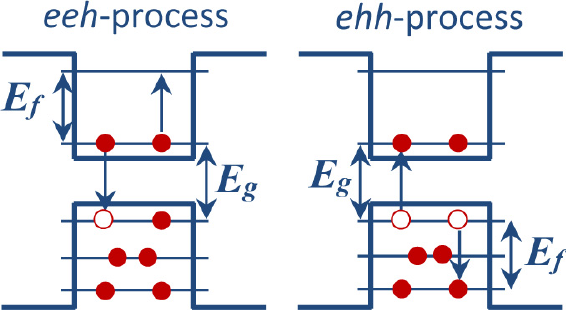}
  \caption{Auger recombination in nanocrystals {via} \textit{eeh} and \textit{ehh} processes. Arrows indicate Auger transitions between the initial and final states.} \label{figure2}
\end{figure}

In order for the Auger process to occur, at least one negative or positive trion (\mbox{exciton + one} electron or exciton + one hole) has to be initially excited in a nanocrystal, as shown in Figure \ref{figure2}. The trion turns into a high-energy single-particle excitation as indicated by the arrows in the figure.

The rate of an Auger-type transition can be written as
\begin{equation}
\tau_{A}^{-1} = \frac{2\Gamma}{\hbar}\sum_f\frac{|U_{if}|^2}{\Gamma^2+(E_f-E_g)^2},
\label{tau_A}
\end{equation}
where $U_{if}$ is the matrix element of the two-particle Coulomb interaction operator, the two-electron wave functions are approximated by products of the single-electron ones and the $\delta$-function expressing the energy conservation in the Fermi's Golden Rule has been broadened with the half-width $\Gamma=10$ meV. All the possible final states of the Auger electron with energy $E_f$ (shown in Figure \ref{figure2}) are summed up. The initial states were chosen so that electrons and holes are in the LUMO and HOMO states, respectively.

Usually, the Auger recombination is an efficient process that can ``shunt'' other photon-assisted processes such as light emission or multiple exciton generation.~For instance, in~Si crystallites, the Auger rates can be sufficiently high, as the measurements~\cite{Stolle,Pevere} and calculations~\cite{JPCC,Sevik,Delerue1,Mahdo,Semi} show. Therefore, suppression of the Auger process is an important practical problem. 

It turns out that coating with halogens can slow down the Auger processes in Si NCs. Below we present the calculated  Auger rates and demonstrate their reduction due to the surface halogenation of Si crystallites. The calculations take into account both the long-range ($U_0({\bf r}_1,{\bf r}_2$)) and the short-range ($U_1({\bf r}_1,{\bf r}_2$)) parts of the carrier--carrier interaction. The first one is a macroscopic electrostatic field modified by the nanocrystal boundary,
\begin{eqnarray}
U_0({\bf r}_1,{\bf r}_2) & = & \frac{e^2}{\varepsilon |{\bf r}_1-{\bf r}_2|}+\frac{e^2(\varepsilon - 1)}{\varepsilon R}
\nonumber \\
& & \times \sum_{l=0}^{\infty}\frac{(l+1)P_{l}(\cos\theta)}{l\varepsilon+l+1}\frac{r_1^{l}r_2^{l}}{R^{2l}}\,,
\end{eqnarray}
while the second one describes at a microscopic level the point charges interaction at short~distances:
\begin{equation}
U_1(r) = \frac{e^2}{r}[Ae^{-\alpha r}+(1-A-1/\varepsilon)e^{-\beta r}]\,.
\end{equation}

Here, $r=|{\bf r}_1-{\bf r}_2|$, $A=1.142$, $\alpha = 0.82/a_B$ and $\beta = 5/a_B$~\cite{PRB}, where $a_B$ stands for the Bohr radius, and we set $\epsilon_1=1$ as before. The calculated values of $\tau_{A}^{-1}$ are listed in {Table} 2 for both \textit{eeh} and \textit{ehh} trion annihilation in the Si$_n$X$_m$ crystallites considered here.

\vspace{0.25 cm}
\noindent {\bf Table 2.} Calculated Auger rates ($\times 10^{10}$ s$^{-1}$) for \textit{eeh} and \textit{ehh} processes in halogenated silicon nanocrystals.

\begin{tabular}[c]{ccccccc}
  \hline
   & X=H & X=Cl & X=Br & X=H & X=Cl & X=Br \\
   & ({\it eeh}) & ({\it eeh}) & ({\it eeh}) & ({\it ehh}) & ({\it ehh}) & ({\it ehh}) \\ \hline
  Si$_{35}$X$_{36}$ & 390 & 3 & 4 & 40 & 163 & 10 \\
  Si$_{59}$X$_{60}$ & 78 & 0.3 & 7 & 5 & 27 & 19 \\
  Si$_{87}$X$_{76}$ & 120 & 4 & 2 & 1780 & 59 & 11 \\
  Si$_{123}$X$_{100}$ & 8 & 3 & 280 & 108 & 25 & 24 \\
  Si$_{147}$X$_{100}$ & 59 & 84 & 2 & 474 & 20 & 65 \\
  Si$_{163}$X$_{116}$ & 120 & 21 & 100 & 47 & 9 & 19 \\
  Si$_{175}$X$_{116}$ & 270 & 2 & 2 & 161 & 90 & 0.8 \\
  Si$_{317}$X$_{172}$ & 0.3 & 0.4 & 0.02 & 5 & 3 & 78 \\ \hline
\end{tabular}

\vspace{0.25 cm}

The surface halogenation of Si nanocrystals also slows down the Auger recombination according to the data presented in {Table} 2. In order to quantitatively analyse the effect of halogenation on the Auger rates, we have calculated the average decimal logarithms of $\tau_{A}^{-1}$ for the Br-, Cl- and H-passivated nanocrystals. In the case of the \textit{eeh} Auger recombination one has obtained: $\langle\lg(\tau_{A(\text{H})}^{-1})\rangle =11.66$, $\langle\lg(\tau_{A(\text{Cl})}^{-1})\rangle =10.55$ and $\langle\lg(\tau_{A(\text{Br})}^{-1})\rangle =10.60$. In the case of the \textit{ehh} Auger recombination the rates also decrease but not so strongly: $\langle\lg(\tau_{A(\text{H})}^{-1})\rangle =11.86$, $\langle\lg(\tau_{A(\text{Cl})}^{-1})\rangle =11.44$ and $\langle\lg(\tau_{A(\text{Br})}^{-1})\rangle =11.20$. It should be noted also that strong reduction of the Auger recombination rates was theoretically shown by Califano~\cite{Calif} for GaSb NCs whose surfaces were passivated with atoms of electronegative elements.

\subsection{Capture on Dangling Bonds}

\begin{figure}[t]
  \includegraphics[scale=0.9]{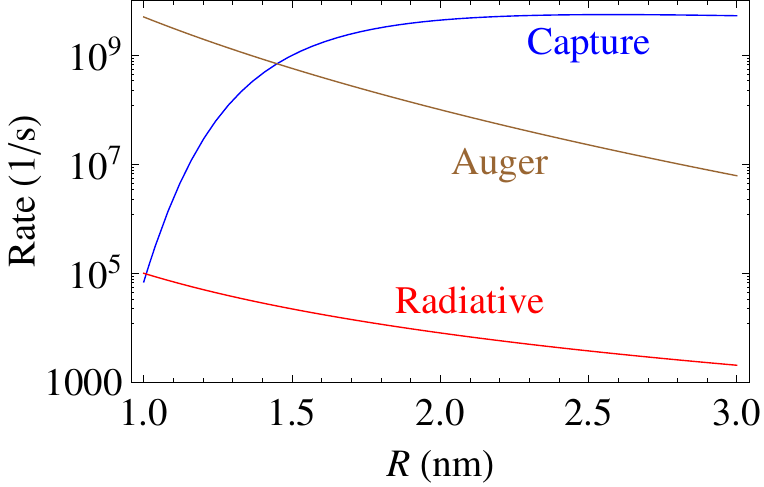}
  \caption{Calculated rates of carrier trapping on dangling bonds compared to the rates of radiative phonon-assisted transitions and Auger recombination in Si nanocrystals.} \label{figure3}
\end{figure}

There is one more relatively fast process inhibiting efficient light emission from NCs, which is photo-carrier trapping on surface defects called P$_{b}$-centres or dangling bonds. The P$_b$-centres produce rather deep energy levels within the nanocrystal band gap, which act as electron traps. The capture on neutral dangling bonds is a multi-phonon process, and its rate strongly depends on temperature. The earlier performed calculations~\cite{book} for Si crystallites yielded $\tau_{C}^{-1}$ shown in Figure \ref{figure3} as function of the nanocrystal radius in comparison with the rates of the Auger recombination and radiative transitions.

It can be seen from Figure \ref{figure3} that for the NC radius greater than $\sim$1 nm, the capture rate becomes larger than the radiative recombination rate. Moreover, starting from $R$ close to 1.5 nm, the trapping on dangling bonds turns out to be even faster than the Auger recombination. Obviously, for nanocrystals whose radii are greater than 1.5 nm, the nonradiative trapping on the surface defects suppresses all the other competing relaxation processes inside the NC if the surface dangling bonds are not passivated.

\subsection{Phonon-Assisted Relaxation of Hot Carriers}

It is well known that highly excited (``hot'') electrons, holes or excitons may relax to the lower states in order to minimize the system energy. In bulk semiconductors and QWS, where the electronic energy spectra are continuous, such a relaxation is accompanied by the emission of phonons caused by the interaction between electrons and lattice vibrations. This interaction also plays a major role in QDs and it is generally accepted that the mechanisms are essentially the same in nanocrystals and bulk materials. 

The most universal mechanism of coupling between the electrons (or holes) and long-wavelength acoustic phonons is through the volume deformation potential. The bottom of a non-degenerate band (e.g.,  conduction band in materials with zinc-blend structure) is shifted proportionally to the (local) relative variation of the volume~\cite{Blacha1984}, 
$$
\hat H_{e-AP} = a_c\sum_{\nu}{\left (\nabla \cdot \hat {\bf u}_{\nu }(\bf r)\right )}\; ,
$$
where  $a_c$ is the bulk deformation potential constant, and $\hat {\bf u}_{\nu }$  denotes the atomic displacement operator corresponding to an acoustic phonon mode $\nu$ (it is expressed in terms of phonon creation and annihilation operators in usual way).
For optical phonons, there are two mechanisms: (i) a universal optical deformation potential (ODP) coupling and (ii) the Fr\"ohlich one characteristic of polar materials. In bulk semiconductors with cubic structure, for long-wavelength optical phonons, the ODP coupling vanishes by symmetry for any non-degenerate band but it is non-zero for holes near the top of the valence band~\cite{Blacha1984}. Its expression can be found in the book~\cite{Rogach2008} (Chapter 8) or in \cite{Hamma2008}.~The Fr\"ohlich interaction Hamiltonian for electrons is simply given by 
$$
\hat H_{e-OP} = -e\sum_{\nu}{\hat \varphi_{\nu }(\bf r)}\; ,
$$  
where $\hat \varphi_{\nu }$ is the electrostatic potential operator corresponding to an optical phonon mode $\nu$. With the known electron and hole wavefunctions and phonon displacements obtained from lattice dynamics equations, the calculation of the coupling matrix elements is straightforward. Even though the Fr\"ohlich interaction usually dominates in QDs made of polar semiconductors, the ODP mechanism is important for some of them (e.g., InP) as revealed by modelling of QD resonant Raman spectra~\cite{Hamma2008}. 

The so-called rigid-ion model is more suitable when describing the electron--phonon interaction in multi-valley semiconductors, such as Si and Ge~\cite{Glemb}, where the processes of inter-valley scattering are often important. 
Within this model, the electron--phonon interaction is described by the following operator (which describes both acoustic and optical phonons),
$$
\hat H_{e-P} = -\sum_{\nu;\,{\bf n},s}{\hat {\bf u}}_{\nu}({\bf R}_{{\bf n}s})\cdot \nabla V_{at}({\bf r}-{\bf R}_{{\bf n}s}).
$$

Here, $V_{at}({\bf r}-{\bf R}_{{\bf n}s})$ stands for the electron potential energy in the field of the $s$-th atom situated in the ${\bf n}$-th primitive cell of the lattice and ${\bf R}_{{\bf n}s}$ is the position-vector of this atom in equilibrium. In practice, atomic pseudopotentials are used.

The discreteness of the energy spectrum in quantum dots implies that the rate of the single-phonon emission should strongly depend on the level spacing~\cite{Inoshita}; in particular, electronic transitions causing the phonon emission should be impossible if the spacing is larger than the phonon energy. This simple idea gave rise to the ``phonon bottleneck'' concept, which predicts the inefficiency of hot carrier relaxation by emission of phonons in QDs~\cite{Bastard}.  However, this prediction relies on the assumption that the phonon emission is irreversible, with a probability described by the perturbation theory (Fermi's Golden Rule), which may not be reliable in QDs. The electron--phonon interaction can be enhanced in nanocrystals because of the spatial confinement of both electrons and phonons and, therefore, multiple scattering processes are important. It means that the electron--phonon interaction in QDs, in principle, {must be treated in a non-perturbative and non-adiabatic way, leading to the energy spectra described by polaronic quasi-particle excitations} ~\cite{Stauber,Jacak,Vasilevskiy2004,Sauvage,Miranda,Calleja,Stauber-Vasilevskiy}. 
In other words, virtual transitions between different electronic levels, assisted by phonons and not requiring energy conservation may be important enough to guarantee a significant modification of the exciton energy spectrum and dynamics~\cite{Rogach2008}. 
Thus, does the phonon bottleneck in QDs really exist? Experimental results still are not completely conclusive. On the one hand, an efficient relaxation of optically created excitons was reported in a number of works studying self-assembled QDs (SAQDs)~\cite{Marcinkevicius,Sun,Muller}, with both PL emission rise time~\cite{Marcinkevicius,Sun} and photo-induced intraband absorption decay time~\cite{Muller} below or of the order of 10 ps. Studies performed on nanocrystal QDs~\cite{KlimovPRL2005,Hendry}, where exciton energy level spacings are larger than in SAQDs, also revealed ultrafast intraband relaxation with a
characteristic time in the picoseconds' domain, which is characteristic of CdSe and CdTe QDs in general~\cite{Rogach2008}. On the other hand, there are published experimental results indicating that the relaxation of optically created excitons can be slow (in the nanoseconds' range), both in self-assembled~\cite{Heitz,Urayama} and nanocrystal~\cite{Sionnest,Nozik2001} QDs.

It is important to realize that the polaron concept by itself cannot explain intraband relaxation of carriers in QDs as polaron is a {\it stationary} state of an electron (or exciton) coupled to optical phonons. Some additional interactions should therefore be responsible for the polaron relaxation~\cite{Stauber-Vasilevskiy}. Several possible mechanisms of hot carrier relaxation in QDs have been proposed:
\begin{enumerate}
\item[(i)] The polaron has a rather short lifetime~\cite{Li} because of the
anharmonic effects that lead to a fast decay of nanocrystal's optical phonons forming the polaron. 

\item[(ii)] {Acoustic phonons can provide the possibility of transitions
between different (exciton-)} polaron states formed mostly by the interaction with optical phonons; the polaron spectrum is discrete but relatively dense owing to the non-adiabaticity of this interaction~\cite{Vasilevskiy2004}. If the acoustic
phonon spectrum is continuous, this additional interaction would drive the polaron dynamics towards equilibrium.

\item[(iii)] In the strong confinement regime where the electron's kinetic energy is larger than the electron--hole interaction, the electron (eventually dressed by phonons and forming the polaron) can relax by an Auger-type mechanism.  The excess
energy is first transferred from the electron to the hole through
their Coulomb interaction and the subsequent hole's cooling occurs via
emission of acoustic phonons~\cite{Kharchenko,Zunger}. It can be feasible because the hole level spacings are relatively small in QDs and match the continuum of acoustic phonon~energies.
\end{enumerate}

There is no final conclusion concerning the relevance of each of the three mechanisms to the breaking of the phonon bottleneck in QDs; however, the first one seems to be the most popular in the literature, even though calculations yielded too low relaxation rates for this mechanism, of the order of 1 ns$^{-1}$, for a typical quantum dot considering anharmonic (Gr\"uneisen) parameters characteristic of bulk materials and the Fr\"ohlich interaction with spherically confined optical phonons and electrons~\cite{Stauber-Vasilevskiy}.
Moreover, the electron--phonon interaction in nanocrystals of non-polar materials (see below) may be too weak to make the polaron effect important. If this is the case, the relaxation proceeds via multi-phonon processes (like temperature-induced interband excitation of charge carriers in bulk semiconductors). Usually, multi-phonon processes are much slower compared to the single-phonon transitions. Therefore, the bottleneck effect does not disappear completely.     

Estimations of the single-phonon relaxation rates in Si nanocrystals~\cite{Prokofiev,YassiAPL} yield typical values of the order of $10^{10}$--$10^{12}$~s$^{-1}$. At the same time, the multi-phonon relaxation rates calculated within the Huang--Rhys model~\cite{book,YassiPE,Moskal} for various transitions vary from $10^7$~s$^{-1}$ to $10^{11}$~s$^{-1}$ for nanocrystals whose diameters do not exceed 5~nm. Note that the rates sharply reduce as the nanocrystal diameter decreases. Similar values of the multi-phonon relaxation rates were obtained experimentally for InGaAs/GaAs SAQDs~\cite{Ohnesorge}. Therefore, the precise reason for very fast exciton dynamics characteristic of CdSe and CdTe colloidal QDs remains an open question, in our view. 

\section{Multiple Exciton Generation}
\label {sec_MEG}

\begin{figure}[t]
  \includegraphics[scale=1]{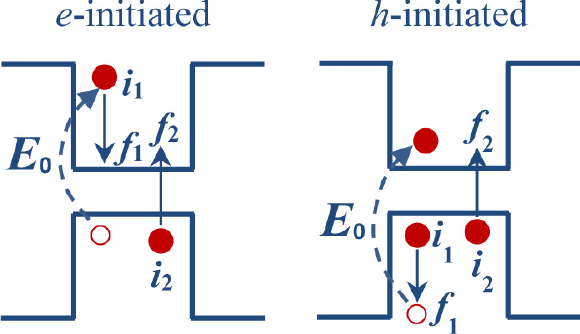}
  \caption{Schematic representation of the electron- (\textbf{left}) and hole-initiated (\textbf{right}) process of multi-exciton generation in a nanocrystal. Initial electron configurations are shown. Dashed arrow indicates highly excited exciton with the excitation energy $E_0$ created by an absorbed photon. Vertical solid arrows indicate electron transitions to the final states. After the transition, the number of excitons increases by one.} \label{figure4}
\end{figure}

As demonstrated above, the surface halogenation suppresses both radiative and Auger recombination in nanocrystals. These processes are reverse ones with respect to the carrier multiplication (or multi-exciton generation): they tend to decrease the number of excitons in a system, while the process of multi-exciton generation, shown schematically in Figure \ref{figure4}, has an opposite trend. Initially, a high-energy photon creates a highly excited electron--hole pair which then reduces its energy creating one more electron-hole pair with lower energy. As a result, two (or even more) excitons can arise in the system after absorption of a single~photon.

The multi-exciton generation is a fundamental process for photovoltaics, where light energy transforms into electric current. Its occurrence in NCs has been experimentally confirmed~\cite{Klimov,Stolle,Beard,Nozik,Boer,Trinh}. In order to be more efficient, this process should be faster than other competitive processes taking place along with exciton generation, such as inter-band radiative recombination or Auger recombination. From this point of view, slowing the latter down caused by surface halogenation is an extremely positive factor. It is important to understand how halogen passivation influences the multi-exciton generation itself.

\begin{figure}[t]
  \includegraphics[scale=0.9]{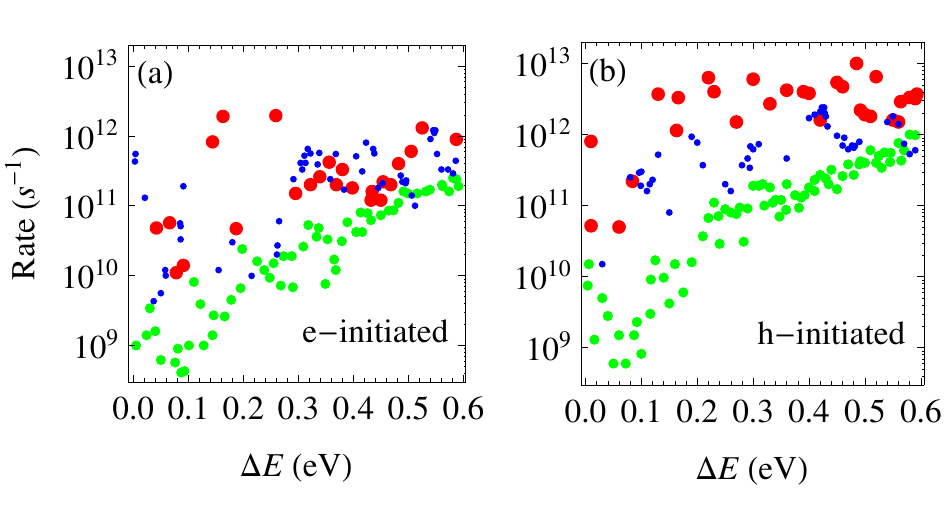}
  \caption{Calculated rates of the exciton generation initiated by highly excited (\textbf{a}) electron and (\textbf{b}) hole (as shown in Figure \ref{figure4}) in Si$_{317}$X$_{172}$ crystallite for X = H (small blue dots), X = Cl (medium green dots) and X = Br (big red dots).} \label{figure5}
\end{figure}

The rates of the multi-exciton generation in the Si$_{317}$X$_{172}$ crystallite (X = H, Cl, and Br) can be calculated using the following relation,
\begin{equation}
\tau_{G}^{-1} = \frac{2\Gamma}{\hbar}\sum_{f_1}\sum_{i_2}\sum_{f_2}\frac{|\langle\Psi_{i_1i_2}|{\hat U}|\Psi_{f_1f_2}\rangle|^2}{\Gamma^2+(E_{i_1} + E_{i_2} - E_{f_1} - E_{f_2})^2},
\end{equation}
where ${\hat U} = {\hat U}_0 + {\hat U}_1$, as before, while $\Psi_{i_1i_2}$ and $\Psi_{f_1f_2}$ are the products of the single-particle Kohn--Sham wave functions of the initial or final electron states participating in the transition, as shown in Figure \ref{figure4}. Here, we present calculated the rates for the Si$_{317}$X$_{172}$ crystallite within the range of excess energy $0<\Delta E <0.6$ eV, where $\Delta E = E_{i_1} - E_{LUMO} - E_g$ for the process initiated by highly excited electron, and $\Delta E = E_{HOMO} - E_{f_1} - E_g$ for the hole-initiated process. The calculated rates are shown in Figure \ref{figure5}. It is evident that the rates rise globally as the excess energy increases, because of the considerable increase in the number of possible states participating in the transitions with increasing $\Delta E$, which opens up many new channels for the realization of exciton generation.

It is important to emphasize that the bromination of a Si crystallite increases the exciton generation rates compared to an H-passivated crystallite, especially if the process is initiated by a highly excited hole. The rates of exciton generation in chlorinated Si crystallite turn out to be lower than those in the hydrogenated one at small excess energies. Meanwhile, upon approaching $\Delta E \sim 0.5$ eV, $\tau_{G}^{-1}$ in the Si$_{317}$Cl$_{172}$ crystallite increases and tends to the typical values observed in Si$_{317}$H$_{172}$ crystallite. Accordingly, it is possible to conclude that halogen coating of Si crystallites, at least, does not reduce their ability to generate excitons, particularly when the excess energies are not too small. This is in contrast with the radiative and Auger recombination processes, where the rates became substantially lower due to the halogenation.

It means that the halogenation of Si NCs can increase the efficiency of the photon-to-exciton conversion, which is defined by an excess of the number of created excitons ($n$) over the number of absorbed photons ($N$)~\cite{Klimov,Haverkort}: $\eta = n/N > 1$ (internal quantum efficiency). There are many experimental works in which external quantum efficiency was measured in Si crystallites~\cite{Stolle,Beard,Nozik,Boer,Trinh}, as well as in crystallites of IV--VI, II--VI or III--V semiconductors~\cite{Klimov,Schaller,Nair,Davis,Hu,Smith}. The authors reported on the observation of multi-exciton generation in the investigated systems.

On the theoretical side, the consideration of the exciton kinetics in the halogen-coated Si crystallites~\cite{PCCP} has revealed strong dependence of $\eta$ on the quantitative relationship between the rates $\tau_G^{-1}$ and $\tau_A^{-1}$. According to the obtained results the decrease in the Auger rate (caused by the halogenation), and its absence in the multi-exciton generation rate, is accompanied by a gradual increase in the quantum efficiency of the order of a few tens of~percentage points.

\section{FRET in Ensembles of Nanocrystals}
\label {sec_FRET}

All the processes considered above may occur in isolated nanocrystals. Meanwhile, usually in experiments, as was already pointed out in the Introduction, one deals with ensembles of NCs, where a non-radiative energy exchange between them takes place and strongly influences the ensemble photoluminescence~\cite{Crooker2002,Franzl2004,Lunz2010,Lunz2011,Yu,Kawazoe,PhysE09,JLumin,JAP}. Such an energy transfer occurs by means of the electron tunnelling if the NCs are connected~\cite{Reich} or through the F\"{o}rster-type migration of excitons~\cite{Forster} if the NCs are separated in space. Below we consider the latter mechanism in some detail.

The F\"{o}rster resonant exciton transfer (FRET) takes place mainly through the dipole--dipole interaction\footnote{It has been shown~\cite{Curutchet2008} that, in contrast to the case of FRET
  between donor and acceptor molecules, where the dipole approximation breaks
  down at lengthscales comparable to molecular dimensions, it works fairly well
  when donor and/or acceptor is a spherical QD, even at contact donor-acceptor
  separations.} of two QDs (a donor and an acceptor). The Quantum Electrodynamics theory of FRET developed in \cite{Andrews1987} reproduces the results obtained in a simpler way by F\"{o}rster~\cite{Forster} and Dexter~\cite{Dexter1953} who considered the electrostatic interaction of two dipoles in two adjacent crystallites~\cite{Allan},
\begin{equation}
V({\bf r}_1,{\bf r}_2) = \frac{\kappa e^2}{\epsilon_{1}b^3}\left[{\bf r}_1\cdot {\bf r}_2 - 3\frac{({\bf r}_1\cdot {\bf b}) ({\bf r}_2\cdot{\bf b})}{b^2} \right]\,,
\label{V}
\end{equation}
where ${\bf b}$ is the inter-crystallite centre-to-centre vector. In one crystallite, the electron--hole pair annihilates and transfers its energy into a neighbouring crystallite where a new electron--hole pair is excited. Thus, virtual transfer of excitons between two crystallites can be realized without charge transfer.

In order to calculate the rate of the F\"{o}rster exciton transition from a QD with a radius $R_1$ into a neighbouring one with a radius $R_2$, we use, as before, the Fermi's Golden Rule:
\begin{equation}
k_{F} = \frac{2\Gamma}{\hbar}\frac{|\langle\Psi_i|{\hat V}|\Psi_f\rangle|^2}{\Gamma^2 + (E_g(R_1) - E_g(R_2))^2}\,.
\label {k_F}
\end{equation}

Here, ${\hat V}$ is the dipole interaction operator identical to (\ref{V}), $\Psi_i = \psi_c({\bf r}_1)\psi_v({\bf r}_2)$ is the wave function of the initial two-particle state with the energy coinciding with the energy gap of the first crystallite, $E_g(R_1)$, and $\Gamma$ is a phenomenological damping parameter. Initially, there is one electron in the conduction band of the first crystallite, with the wave function $\psi_c({\bf r}_1)$, while a hole exists in the valence band. In the second QD, the electron occupies a valence band state described by the wave function $\psi_v({\bf r}_2)$. In the final state, the system has the wave function $\Psi_f = \psi_v({\bf r}_1)\psi_c({\bf r}_2)$, corresponding to the electron-hole pair transferred to the second QD.

The result is essentially the same as if one considered the interaction between two transient point dipoles, $\bf d_D$ and $\bf d_A$ (located at points $\bf r_1$ and $\bf r_2$) assuming that they are stationary. The square of the matrix element in (\ref {k_F}) (the donor--acceptor coupling parameter) can be written as~\cite{Stavola1985} 
\begin{equation}
J^{2} = \nu \frac{\kappa^{2}{d_{D}^{2}d_{A}^{2}}}{\epsilon_{1}^{2}b^{6}}\, , 
\label{ghj}
\end{equation}
where $\nu $ is a factor of the order of unity that takes into account the relative orientation of the dipoles~\cite{Lyo}, $\nu =\frac{2}{3}$ if these orientations are completely random.

The broadening of the exciton transfer resonance, $\Gamma$ can be related to the spectral overlap of the emission and absorption spectra of the donor and acceptor, respectively~\cite{Dexter1953}, and the transfer rate can be expressed as~\cite{mohseni2014quantum}
\begin{equation}
\frac {k_{F}}{\gamma _0^{(D)}} = \frac{3\kappa^{2}c^{4}}{4\pi \epsilon _1^{2}b^6}Q_A
\int_{-\infty}^{+\infty} \frac {d\omega}{\omega ^4} I_{A}(\omega) L_{D}(\omega)\, , 
\label{forster2}
\end{equation}
where $\gamma _0^{(D)}$ is the spontaneous emission rate of the donor,  $I_{A}(\omega )$ is the absorption lineshape function of the acceptor, $L_{D}(\omega)$ is the emission lineshape function of the donor (both are normalized to unity) and $Q_A$ is the frequency-integrated absorption cross section of the acceptor QD,
\begin{equation}
Q_A = \left (\frac{\pi\hbar c}{E_g(R_2)}\right )^2 \gamma _0^{(A)}
\label{Q_A}
\end{equation}
with $\gamma _0^{(A)}$ being the spontaneous emission rate of the acceptor. Notice that the depolarization factor $\kappa^{2}${,} Equation (\ref {kappa}{)}, which distinguishes QD donors and acceptors from, e.g., molecules, has been included explicitly in Equation (\ref {forster2}).

The FRET rate decreases rather quickly with the donor-acceptor distance, $k_{F}=\gamma _0^{(D)} (\frac {b_0}{b})^{6}$, where $b_0$ is a parameter of the dimension of length named ``F\"orster radius'' whose definition is clear from Equation (\ref{forster2}).

In ensembles formed of direct-band-gap II--VI or III--V semiconductor crystallites (QDs), the exciton transfer has the rates $\sim$$10^8$--$10^{9}$$\cdot$s$^{-1}$, as measurements~\cite{Kagan96,Crooker2002,Lunz2010,Lunz2011} and calculations~\cite{Allan,Scholes,Lyo} show. Notice that $k_F$ depends on $b$ and the above values probably correspond to somewhat different donor--acceptor distances. Experimental observations of the exciton transfer were carried out with ensembles of closely packed monodisperse CdSe nanocrystals as well as of two-size three-dimensional mixtures~\cite{Lunz2010} and bilayered CdSe nanocrystal systems~\cite{Lunz2011}. It was shown that the inter-layer transfer in bilayered ensembles with controlled donor-acceptor separation turns out to be more efficient than in the 3D ensembles of the monodisperse and the two-size-mixed crystallites~\cite{Lunz2011}. Note that the measured and computed FRET rates are of the same order of magnitude as the radiative recombination rates in high-density ensembles of colloidal II--VI crystallites. Moreover,~a few measurements of the F\"orster radius have been performed by controlling the distance between two different groups of QDs of different size, acting as donors and acceptors~\cite{Shafiei2011}, or by preparing a homogeneously blended solid-state films~\cite{Mork2014} composed of two groups of dots. In both works, CdSe/ZnS core--shell QDs were used and the reported results for the F\"orster radius are 14--22 nm and 8--9 nm, respectively; notice that the former work~\cite{Shafiei2011} used larger nanocrystals. Despite the uncertainty in experimental conditions and difficulty to evaluate the spectral overlap in Equation (\ref{forster2}) for {\it individual } QDs, there is a consensus that  $R_F$ typically is of the order of 10 nm for highly luminescent colloidal nanocrystals.   

{In ensembles of silicon crystallites the exciton transfer turns out to be much slower than the radiative recombination, its rates are two to three orders of magnitude lower~\cite{Allan,JCTN,Gusev}} (smaller than $\sim$$10^3$$\cdot$s$^{-1}$). Doping of silicon nanocrystals with phosphorus allows to increase the rates up to the values comparable with those of radiative recombination~\cite{PRB2} ($\lesssim 10^7$~s$^{-1}$ for nanocrystal radius $R\gtrsim 1$~nm). Nevertheless, these values remain still smaller than the FRET rates for colloidal QDs made of direct-gap semiconductors.

For the sake of completeness, F\"orster-type resonant energy transfer from a QD exciton can occur also when the final elementary excitations have a different nature, for instance, they are phonons or molecular vibrations. The latter can involve ligand molecules present in the vicinity of a colloidal QD or solvent molecules and is particularly important for quantum dots emitting in the infrared~\cite {Aharoni2010,Qiannan2016}.
Such a process was termed electronic-to-vibrational energy transfer (EVET)~\cite{Aharoni2010}. The importance of EVET was demonstrated by comparing the luminescence properties of HgTe QDs dissolved in two chemically identical solvents: H$_2$O and D$_2$O~\cite {Qiannan2016}.

\section{QD Emitters Near a Flat Interface}
\label {sec_env}

Now, we shall discuss the influence of a flat interface between two media  on the emission and FRET rates for a QD emitter located at a certain distance from it.
If the characteristic distances (such as that between the emitter and the interface) are small compared to the EM wavelength, one can treat the problem in the electrostatic approximation where one neglects both the retardation effects and the magnetic field associated with the electric field present in the media; then, the image dipole method can be used to take into account the effect of surface polarization induced by the dipole~\cite{Santos_PRB2014}.
A more general approach consists in using the dyadic Green's function formalism~\cite{AS2017}. If both dielectrics are dispersionless, the spontaneous decay and FRET are affected through the photonic density of states (DOS) renormalization due to the reflection of electromagnetic (EM) waves at the interface.
However, even in this geometrically simplest situation, there is a less trivial effect produced by EM waves created by the polarization in the second medium, totally reflected at the interface so that only an exponentially decreasing field amplitude of theirs reaches the dipole. Yet, this effect can be dominating at small distances, not only for radiative decay~\cite{Lukosz} but also for FRET.
When the second medium is a metal, several contributions to both decay and transfer processes arise that can be associated with (i)~propagating EM waves (so-called radiative losses), affected by the presence of the second medium; (ii) coupling to propagating surface plasmons (SPs); and (iii) Ohmic losses (when the exciton energy is irreversibly transferred to heat via electron scattering in the metal).

Even though the subject has been studied during several decades~\cite{Lukosz,Drexhage,Chance} and is described in topical reviews~\cite{Ford-Weber} and books~\cite{Hecht-Novotny}, we find some significant effects that were overlooked in previously published discussions and, in our opinion, are important for potential applications of FRET in photonics and energy harvesting. In particular, we will show that (1) if a QD embedded in a dielectric matrix is located near an interface with another dielectric (a substrate) having a lower dielectric constant, its emission is polarized parallel to the interface; (2) in the opposite case of substrate's dielectric constant higher than that of the matrix, the emission and transfer rates are strongly enhanced (without dissipation) near the interface; and (3) the resonant coupling between SPs propagating along a metal/dielectric interface and excitons confined in QDs located at a distance of the order of the light wavelength from the interface, can be used for long-range FRET.
We present these rates calculated for typical colloidal nanocrystal QDs.

\subsection{Radiative Lifetime Near Interface}
\label {subsec_interface}

We shall consider the QD emitter as a point dipole.\footnote{See~\cite{Gordon2014} for discussion of possible effects beyond this
  approximation.} The radiative decay rate in the presence of other bodies is determined by the total field that acts on the emitter (created by itself and scattered by the bodies) and given by \cite{Hecht-Novotny}
\begin{equation}
\gamma =\mbox {Im} \left [\boldsymbol d^\star \cdot  \left (\mathbf  E_0 (\mathbf r) + \mathbf  E_s (\mathbf r) \right ) \right ]/2\hbar \;,
\label{eq:gamma}
\end{equation}
where $\mathbf {E}_0$ is the dipole field in infinite space.
Here, $\boldsymbol {d^\star}$ is the complex-conjugate of the classical dipole moment, the necessary adaptation for the quantum mechanical transition dipole matrix element consists in multiplying it by 2 in the final result, $\boldsymbol {d} \rightarrow 2\mathbf {d}$~\cite{Hecht-Novotny}.   
The~electric component of the scattered field created by an emitting dipole located in the origin of the reference frame can be written in terms of the Green's dyadic~\cite{Hecht-Novotny}, which is specific of system's geometry:
\begin{equation}
\mathbf E_s (\mathbf r^\prime) =4\pi k_1^2 \mathbf G (\mathbf r^\prime,\mathbf 0)\cdot \boldsymbol d \, .
\label{eq:e-field}
\end{equation}

The $3\times 3$ dyadic matrix $\mathbf G (\mathbf r^\prime,\mathbf r)$ is determined by the Fresnel coefficients, $r^{(p)}$ and $r^{(s)}$, and its explicit expression can be found in ~\cite{AS2017}.

\begin{figure}[t]
\includegraphics*[width=7.5cm]{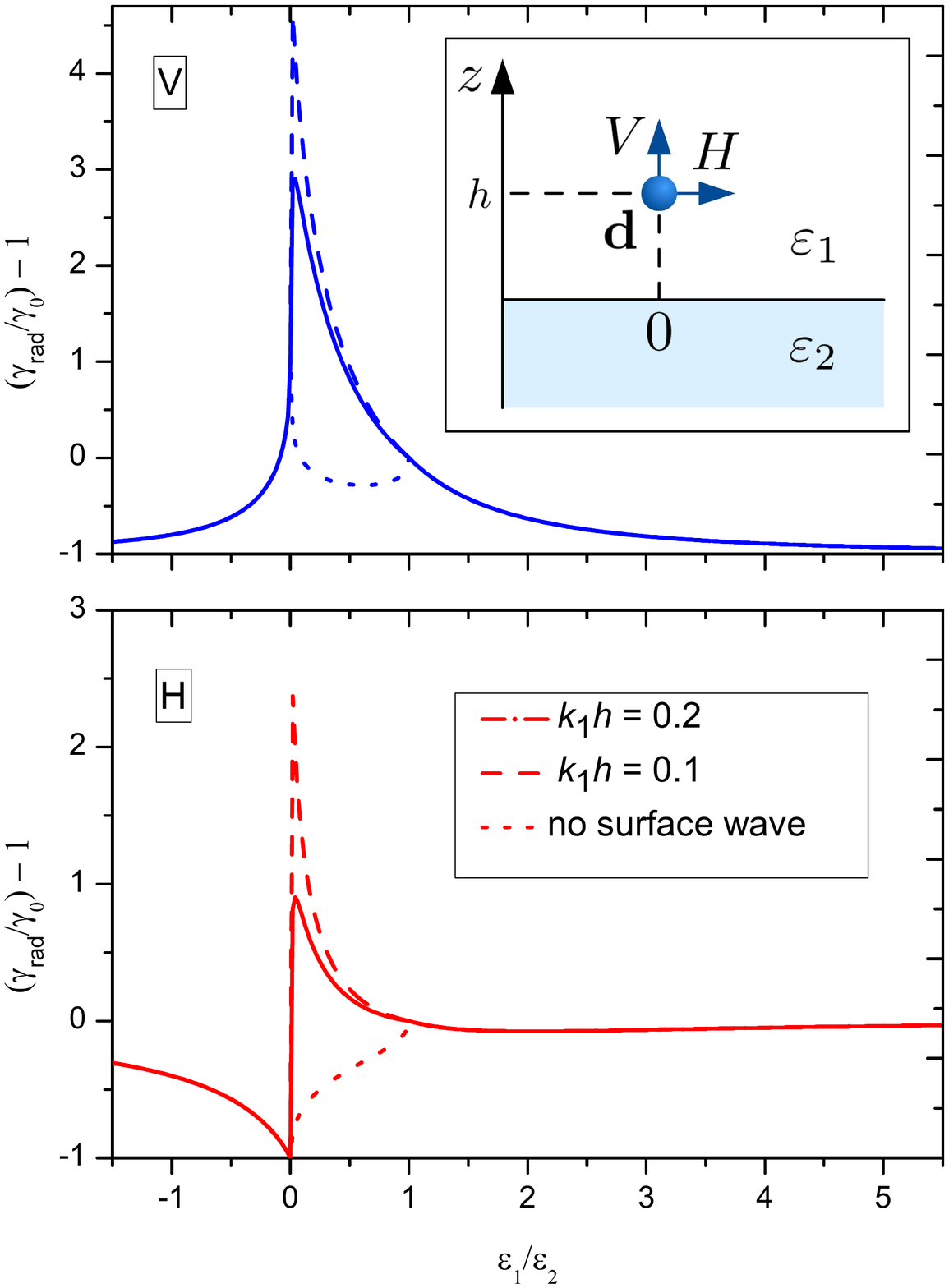}
\caption{{Dependence} of the radiative decay rate, $\gamma $, on the dielectric constant ratio $\varepsilon_1/\varepsilon_2$ for two two different distances to the interface between them, $k_1h= 0.1$ and 0.2. The dotted shows the dependence when the contribution of evanescent waves (that exists for $0<\varepsilon_1/\varepsilon_2<1$) is neglected; it is independent of $h$.
The inset shows two orthogonal orientations of the dipole moment. The upper plot is for the vertical dipole and the lower for the horizontal orientation.} 
\label{fig:radiative}
\end{figure}

{From Equation (\ref {eq:gamma}), one obtains for a dipole perpendicular to the interface (i.e., ``vertical''}, see inset in Figure \ref{fig:radiative}):
\begin{equation}
\frac {\gamma ^{(V)}} {\gamma _0} =1+\frac 3 2 \mbox {Re} \int _0 ^\infty {\frac {s^3} {\sqrt {1-s^2}}r^{(p)}\exp {(2ik_1h\sqrt {1-s^2})}ds} \;,
\label{eq:gamma-V}
\end{equation}
and for a dipole parallel to the interface (i.e., ``horizontal''):
\begin{eqnarray}
\nonumber
& &\frac {\gamma ^{(H)}} {\gamma _0} =1+\frac 3 4 \mbox {Re} \int _0 ^\infty \frac s {\sqrt {1-s^2}}\left [r^{(s)}-(1-s^2)r^{(p)}\right ]\\
& &\times \exp {(2ik_1h\sqrt {1-s^2})}ds \,.
\label{eq:gamma-H}
\end{eqnarray}

Here, $ \gamma _0 $ is the decay rate in an infinite medium with dielectric constant $\varepsilon _1$, given by Equation (\ref{eq_gamma0}). The integrals in Equations (\ref {eq:gamma-V}) and (\ref {eq:gamma-H}) with respect to the normalized in-plane wavevector, $s=q/k_1$, can be divided into two parts, one from 0 to 1 corresponding to propagating waves in the upper half-space and the other from 1 to $\infty $ representing evanescent waves with imaginary wavevector component along $z$ axis perpendicular to the interface, $k_{1z}$.
The latter type of waves exist in two cases: (i) if $\varepsilon_1 < \varepsilon_2$ and both dielectric constants are positive and (ii) if $ \varepsilon_2^\prime \equiv \mbox {Re} (\varepsilon_2) <0$, i.e., when the second medium is a metal.

In the case of two non-dispersive dielectrics,  $ \gamma $ given by Equations (\ref {eq:gamma-V}) and (\ref {eq:gamma-H}) represents only radiative losses of dipole's energy, renormalized by the back action of the scattered waves.
Yet, for $\varepsilon_1 < \varepsilon_2$ there are EM waves (excited by the dipole) propagating in the lower half-space (with real $z$-component of the wavevector inside medium 2, $k_{2z}$) and experiencing total internal reflection at the interface with the upper medium, possessing imaginary $k_{1z}$. The Fresnel coefficient  $r^{(p)}$ has an imaginary part associated with these waves. If the interaction of the dipole with these evanescent waves is neglected (dotted lines in Figure \ref{fig:radiative}), i.e., we set the integral from 1 to $\infty $ in Equations (\ref {eq:gamma-V}) and (\ref {eq:gamma-H}) equal to zero, then $ \gamma $ becomes independent of dipole's distance from the interface.

As can be seen from this figure, the emission decay rate is strongly enhanced, $ \gamma >> \gamma _0 $, at small distances ($k_1h<1$) for both dipole orientations because of the strong interaction with the evanescent waves.
It has been pointed out for plane interfaces by Lukosz and Kunz~\cite {Lukosz}, and it is similar to the coupling to the whispering gallery modes (a kind of surface EM waves) observed for emitters placed inside a micrometer-size dielectric sphere~\cite {Benner1980,Schneipp2002}. As we will show in the next section, this coupling can be used for FRET~enhancement.

%
%

For $\varepsilon_1 > \varepsilon_2$ only propagating waves in the upper medium (with $k_{1z}$ real) are present and $\gamma $ does not depend on $h$.
As can be seen from Figure \ref{fig:radiative}, the emission rate of a dipole oriented perpendicular to the interface is suppressed ($\gamma \rightarrow 0$), while that parallel to the interface has $\gamma $ close to its value in an infinite dielectric, $\gamma _0$. As known, the emission of a spherical QD made of a semiconductor material with cubic crystal lattice does not have any preferential polarization in empty space (if excited by non-polarized light). If the dot is embedded in a matrix with sufficiently high $\varepsilon _1$ and located near the surface, one can expect its luminescence to be strongly polarized.
Note that this is not a quenching effect because there is no dissipation in the system (both $\varepsilon_1$ and $\varepsilon_2$ are real).

\subsection{Non-Radiative Losses to a Metal Substrate}
\label {subsec_metal}

If the substrate is lossy, e.g., a metal, several additional decay channels arise. Mathematically, they are related to the imaginary part of the Fresnel coefficients. Direct evaluation of the integrals in  Equations (\ref {eq:gamma-V})  and  (\ref {eq:gamma-H}) yields the overall effect of all these channels.
However, their relative contributions depend on the distance and the real and imaginary parts of $\varepsilon_2$.
At very small distances ($k_1h<<1$), the dominating contribution is due to Ohmic losses \cite{Chance,Ford-Weber},
\begin{equation}
\frac {\gamma ^{(V,\:H)}_{\mbox{Ol}}} {\gamma _0} =\frac 3 {(2k_1h)^3} \mbox {Im}\left (\frac {\varepsilon_2 - \varepsilon_1}{\varepsilon_2 + \varepsilon_1} \right )\; .
\end{equation}

The coupling to propagating surface plasmons mathematically is described by the pole of $r^{(p)}$, which yields an additional imaginary part of the integral from 1 to $\infty $ in \mbox{Equations~(\ref {eq:gamma-V}) and (\ref {eq:gamma-H})}.
Indeed, the equation $(r^{(p)})^{-1}=0$ determines the dispersion relation of $p$-polarized SPs. The other Fresnel coefficient, $r^{(s)}$, has no poles, and accordingly there are no  $s$-polarized SPs at a metal--dielectric interface.
If we neglect the imaginary part of $\varepsilon_2$, the SP wavevector is real. The resonant coupling occurs for
$$
q_{\mbox{sp}}=\frac {\omega} c \sqrt {\frac {\varepsilon_2^{\prime }   \varepsilon_1}{\varepsilon_2^{\prime } + \varepsilon_1}}
$$
(where $\varepsilon_2^\prime  \varepsilon_1<0$ and $\varepsilon_2^\prime  +\varepsilon_1<0$). The pole contribution to the integral from 1 to $\infty $ in Equations (\ref{eq:gamma-V}) and (\ref{eq:gamma-H}) is calculated explicitly:
\begin{eqnarray}
& &\frac {\gamma ^{(V)}_{\mbox{sp}}} {\gamma _0} =3\pi s_{\mbox{sp}}^5 \frac {\sqrt {-u}} {1-u} \exp {\left (-\frac {2k_1hs_{\mbox{sp}}}{\sqrt {-u}}\right )} \,,\\
\label{eq:gamma-SP-V}
& &\frac {\gamma ^{(H)}_{\mbox{sp}}} {\gamma _0} =\frac {3\pi} 2 s_{\mbox{sp}}^5 \frac {1} {\sqrt {-u}(1-u)} \exp {\left (-\frac {2k_1hs_{\mbox{sp}}}{\sqrt {-u}}\right )} \,,
\label{eq:gamma-SP-H}
\end{eqnarray}
where $u=\varepsilon_2^\prime /\varepsilon_1$ and $s_{\mbox{sp}}=q_{\mbox{sp}}/k_1$.

\begin{figure}[t]
\includegraphics*[width=7.5cm]{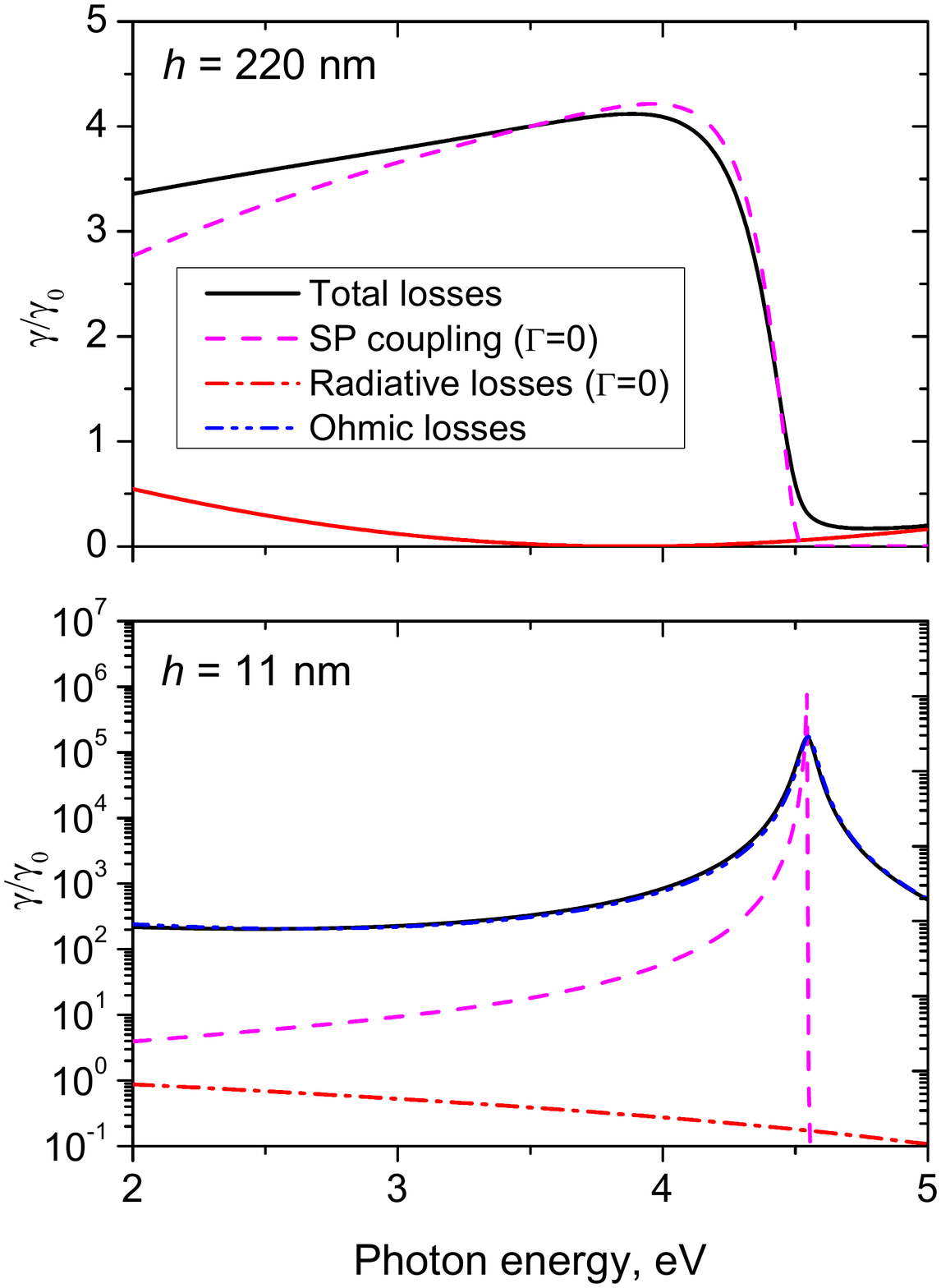}
\caption{{Spectral} dependence of different contributions to the emission decay rate for a vertically oriented emitter located in a dielectric with  $\varepsilon_1 =2$ at two different distances (indicated on the plots) above a flat interface with gold. Note that the contribution of the Ohmic losses (the main cause of the emission quenching near metal surfaces) is negligible in the upper plot, while it is completely dominating in the lower panel. For the calculation of the radiative losses and coupling to surface plasmons, the damping parameter, $\Gamma$ of the Drude model was set equal to zero. }
\label{fig:gamma_metal} 
\end{figure}

Taking $\varepsilon_2 (\omega)$ within the simple Drude model for bulk gold, we calculated $\gamma _{\mbox{Ol}}$,  $\gamma _{\mbox{sp}}$, and the radiative decay rate (neglecting $\varepsilon_2^{\prime \prime}$). They are presented in Figure \ref{fig:gamma_metal}. As expected, at small distances the Ohmic losses dominate (see Figure \ref{fig:gamma_metal}, lower panel). However, as the distance to the interface increases, for $k_1h\sim 1$, the coupling to SPs becomes the main mechanism that determines the lifetime (see Figure \ref{fig:gamma_metal}, upper panel). Although this effect decreases with the distance exponentially,  this decrease can be rather slow, as $\vert \varepsilon_2^\prime \vert /\varepsilon_1>>1$ and $s_{\mbox{sp}}\sim 1$, so it is nearly constant over tens of nanometres. Of course, at still large distances $\gamma _{\mbox{tot}}\rightarrow \gamma _0$.
%
%

Thus, for intermediate distances of the order of a hundred nanometres the radiative decay is mostly due to the coupling to SPs within a broad spectral range (approximately from 2.5 to 4.5 eV). In this range, relevant for colloidal QDs, one can expect also plasmon-enhanced FRET. Even though the enhancement is moderate, the coupling to SPs is not dissipative unless we approach the electrostatic SP resonance frequency, $\omega _{spr}=\omega _p/\sqrt {\varepsilon_2^\infty +\varepsilon_1}$ (here, $\omega _p$ is the bulk plasma frquency and $\varepsilon_2^\infty$ is the background dielectric constant of the metal). In other words, excitation can be transferred from the QD to SPs and back many times without dissipation. It can lead to strong coupling regime characteristic of plasmonic microcavities~\cite{Giannini2011,Dovzhenko2018}.

\subsection{Energy Transfer to a 2D Material}

Graphene, an atomic-thick monolayer of carbon, is a semimetal (or gapless semiconductor) with unusual electronic properties first demonstrated by Geim and Novoselov~\cite{novoselov2004electric}. Low-energy excitations in graphene are massless, chiral, Dirac fermions.~In~neutral graphene, the chemical
potential (hereafter called Fermi level) crosses exactly the Dirac point~\cite{rmp}.
Pristine graphene is transparent in a broad spectral range from the infrared  (IR) to the ultraviolet (UV), with the residual absorption of $\approx$2.3\% originating from interband transitions~\cite{Nair1308}.~The Fermi level, $E_F$, can be shifted by up to $\approx$1 eV with respect to the Dirac point by applying an appropriate gate voltage. The usual doping via adsorbed impurities is also possible~\cite{rmp}. One can say that graphene is a transparent conductor with a tunable conductivity, both static and frequency-dependent (for the frequencies up to the mid-IR)~\cite{Li2008}. It supports propagating surface plasmons whose dispersion relation can be controlled via gate voltage~\cite{c:primer}.
Interestingly, a~specific type of quantum dots can form in graphene, namely, mass profile QDs, a system within current experimental reach~\cite{Galera2014}, for which FRET can also be important~\cite{Stauber2015}.   

Numerous experiments have shown that the nanoparticle or dye molecule emission is quenched in a broad spectral range by a single graphene sheet~\cite{Chen2010,Gaudreau2013,Federspiel2015_QD-G,Goncalves_2016,Raja_nanolett2016}.
It is understandable for the lower-energy part of the spectrum, as graphene is a conductor, although with a relatively low free carrier concentrations (which can reach $10^{13}$ cm$^{-2}$), with the plasma frequency in the THz-to-IR range; it also is supported by calculations. However, the situation is more complex than in the case of a point emitter in the vicinity of a usual~metal.

\begin{figure}[t]
\includegraphics*[width=8.5cm]{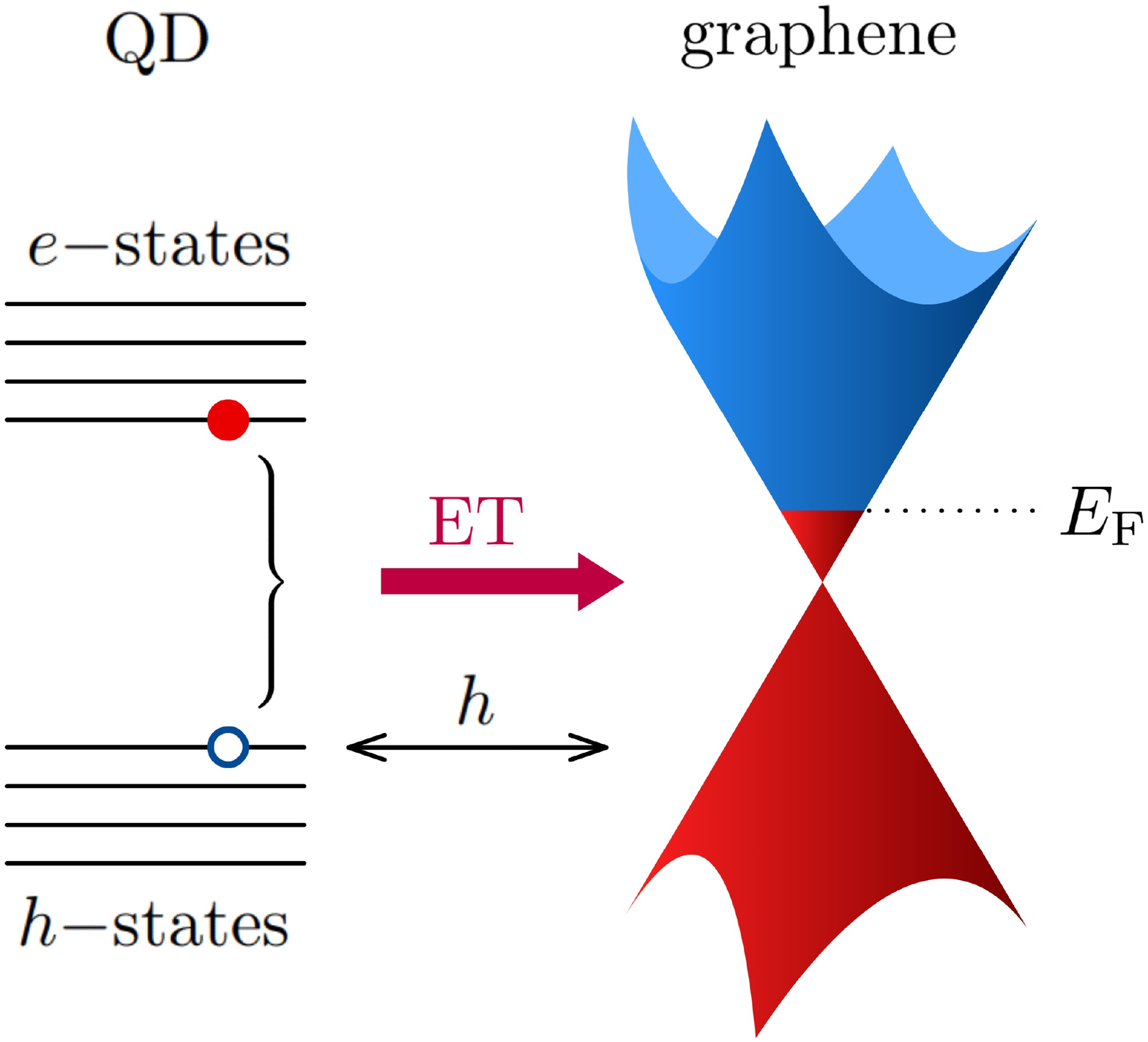}

\caption{Schematics of the exciton energy transfer from a QD to graphene. The left part represents an excited QD with an electron--hole pair. It can be transferred as a whole to the graphene represented by its conduction and valence bands' cones touching in the Dirac point. For the first-order vertical interband transition in graphene, at low temperatures the transfer is possible only if the exciton energy exceeds 2$E_F$.}
\label{fig:dirac_point}
\end{figure}

In addition to the Ohmic losses and energy transfer to surface plasmons, there is an additional mechanism that may be named exciton transfer (ET) to graphene. 
It is similar to FRET and consists in energy transfer to electron--hole pairs generated by inter-band transitions, symmetric with respect to the Dirac point (see Figure \ref {fig:dirac_point}). 
Doped graphene absorbs EM radiation due to inter-band transitions only for energies above $2E_F$, therefore, it also applies to the ET.
The main contribution of this mechanism scales with distance as $h^{-4}$. It can be demonstrated in a rather simple way that elucidates its connection to FRET. 

Let us assume that acceptor dipoles are distributed within a plane (representing the
graphene sheet), then the ET rate can be obtained by integrating (\ref{forster2}) over this plane and written as follows,
\begin{equation}
\frac {k_{ET}(h)}{\gamma _0^{(QD)}} = \rho_A \int _0^{2\pi}d\phi
\int_{0}^{\infty} \frac {r dr b_0^6}{(r^2+h^2)^3} \, , 
\nonumber
\end{equation}
where $\rho_A $ is the surface density of acceptor dipoles. The emitter lineshape may be taken as a $\delta $-function. As $\rho_A Q_A I_A = \beta (\omega)$ is the absorbance of the graphene, we obtain
\begin{equation}
\frac {k_{ET}(h,\omega)}{\gamma _0^{(QD)}} = \frac {3c^4}{8\epsilon _1^2 \omega^4 h^4 }\beta (\omega) \, . 
\label{ET_rate}
\end{equation}

This formula can be derived in a more rigorous way \cite{Gomez-Santos_PRB2011}, with just slightly different numerical coefficients.

%
%

The $h^{-4}$ scaling law was verified in several works and it seems to be obeyed for molecular emitters (down to $h\approx 5\, \text {nm}$)~\cite{Gaudreau2013} but not so much for QDs where deviations were found for small distances~\cite{Federspiel2015_QD-G}. The above theory does not take into account the finite size of the QDs. As a first approximation, one can assume that the point dipole is located at the centre of the QD. In this case, the distance of the dipole from the graphene sheet can be approximated as $h^* = (h+ R_{QD} +s)$ where $R_{QD}$ is the radius of the nanocrystal core of the QD and $s$ is thickness of the organic shell for a colloidal QD. The latter can be estimated as the chain length of the capping agent, $s\sim 1.5\, \text {nm}$~\cite{Rogach2008}. Using this correction, $h\rightarrow h^*$, indeed improves the agreement between the theory and experiment ~\cite{Federspiel2015_QD-G}.
However, one needs to bear in mind also the possibility of charge transfer from an excited QD to graphene if the distance is sufficiently small for tunnelling. Indeed, the demonstration of hybrid graphene-QD phototransistors~\cite{Konstantatos2012} and solar cells~\cite{Sun2011} where the light is absorbed in the QDs and changes the electric current in graphene implies that such processes can take place in specially designed structures. 

Calculations including both intra-band (i.e., Drude plasmons) and inter-band transitions in graphene show that there can be a no-quenching spectral window for an emitter over a strongly doped graphene sheet~\cite{Koppens2011}. Indeed, the graphene absorbance, $\beta (\omega)$, due to Drude plasmons decreases with the frequency, while the inter-band transitions contribute to $\beta$ (and, therefore, to the quenching) only for $\omega >2E_F/\hbar$~\cite{c:primer}. As the Fermi level in graphene can be tuned electrically, it opens an interesting possibility of electrical switching of the QD emission by controlling the quenching rate; it has been demonstrated experimentally in ~\cite{Lee_nanoletters2014,Salihoglu2016}.

Energy transfer (ET) from QDs to other 2D materials, namely, few-atomic-layer-thick semiconductors (transition metal dichalcogenides (TMDs)) has also been investigated in the recent years~\cite{Prasal_nanolett2015,Raja_nanolett2016,Prins2014,Zang2016,Karanikolas2016}. These materials support robust excitons, which determine their optical properties in the visible range (along with the inter-band transitions) and they depend strongly on the number of monolayers in the material~\cite{Wang2018}.   
The ET from a QD exciton to the continuum of states above the band gap in the 2D material should work similar to the case of graphene (Equation (\ref{ET_rate})). However, it was found that the donor QD emission is quenched most strongly near monolayer MoS$_2$ and then the effect decreases with the number of monolayers (presumably keeping the QD distance to the semiconductor surface fixed), in contrast with the case of few-layer graphene~\cite{Raja_nanolett2016,Prins2014}. Meanwhile, the tendency similar to graphene (quenching increases with the number of monolayers) was found for the less studied TMD SnS$_2$~\cite{Zang2016}. On the theoretical side, it has been suggested that the total spontaneous emission rate of a quantum emitter can be enhanced several orders of magnitude due to the excitation of surface exciton--polariton modes supported by the 2D semiconductor~\cite{Karanikolas2016}. In the vicinity of an excitonic transition, the real part of the dielectric function of the semiconductor material is negative, like in a metal, so~one can indeed expect an enhancement of the local photonic DOS and, consequently, the~Purcell effect~\cite{Gomes2020}. However, it would require matching of the energies of the QD and 2D excitons.       

\subsection{FRET between QDs Near Interface}

\begin{figure}[t]
\includegraphics*[width=8.5cm]{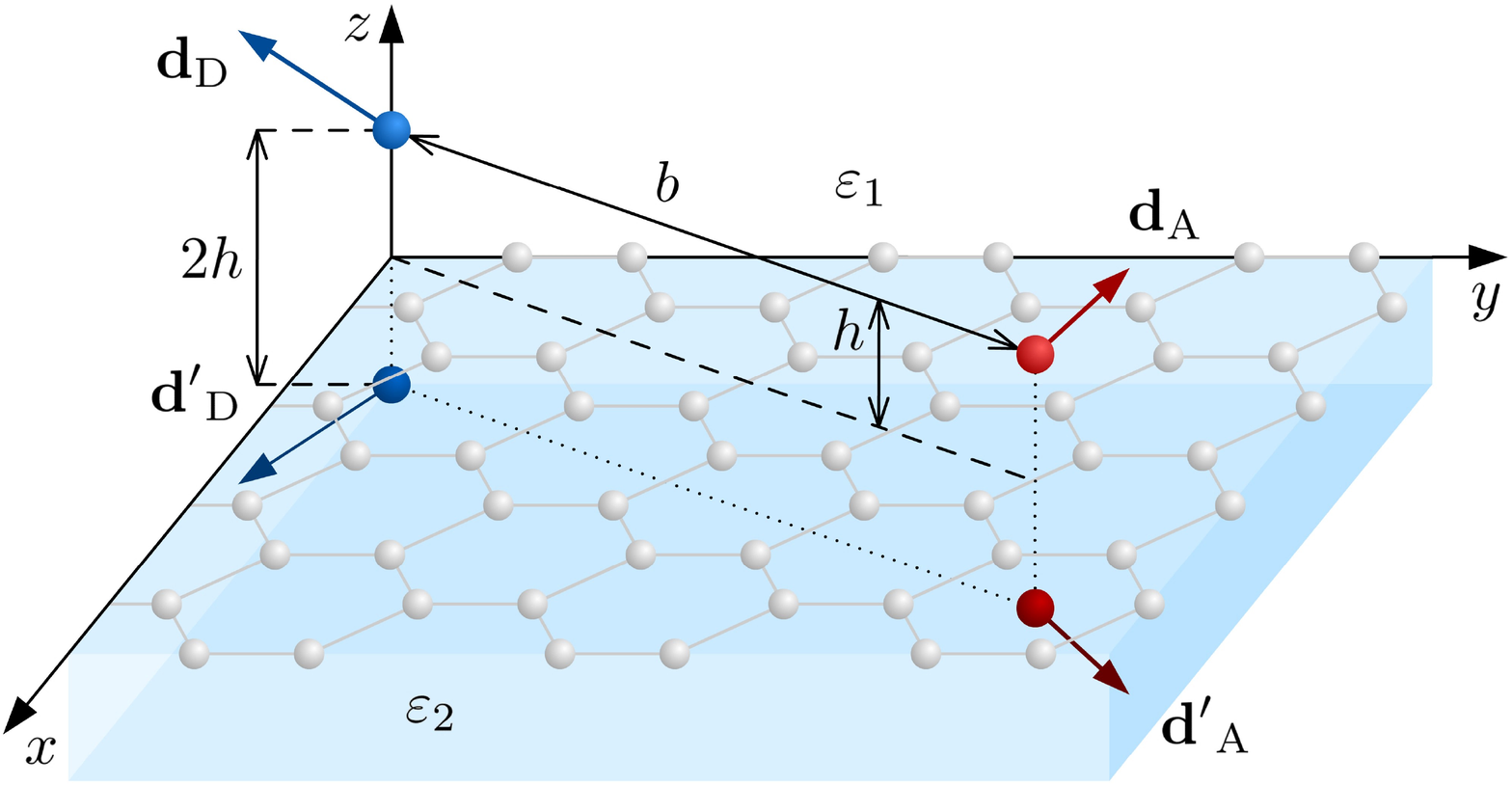}
\caption{Schematics of two QDs on top of a graphene sheet, interacting via their real transient dipoles and image dipoles representing transient charge distributions in the graphene sheet.}
\label{fig:two_dipoles_on_graphene}
\end{figure}

In the vicinity of a polarisable material, such as graphene, the dipole--dipole interaction responsible for the FRET between two QDs becomes renormalized because of the additional interaction between the induced charges. In the electrostatic limit, these additional interactions can be modelled with image dipoles (see Figure \ref {fig:two_dipoles_on_graphene}).~This physical mechanism, of course, is not unique to graphene and can take place in the vicinity of plasmonic nanoparticles~\cite{Lunz2012,Knorr2013,Komarala2008}. The effect of plasmonic enhancement of FRET has been detected experimentally by observing a decrease of the emission decay rate of donor QDs and a corresponding increase of the luminescence intensity for acceptor dots when they are placed in the vicinity of gold nanoparticles~\cite{Lunz2012}. Of course, {if the concentration of the nanoparticles becomes too high, the emission is quenched for both donor and acceptor QDs because of the energy transfer to lossy modes in the metal (the Ohmic losses, Section~\ref {subsec_metal}).}

%
%

Graphene is a conductor with electrically controllable conductivity and, in principle, it promises a possibility of controlling the inter-dot FRET rate. The exciton energy transfer between two QDs located near a graphene-covered interface occurs not just through the direct dipole--dipole interaction but also through polarization charges that they induce on the interface.
One can think that the image dipoles of Figure \ref {fig:two_dipoles_on_graphene} have magnitudes that depend on the optical conductivity of graphene and, therefore, on its Fermi energy~\cite{Santos_PRB2014}.  
This interaction can be written in the form 
$$V_{FRET}=\frac 1{\epsilon_1}\bf d_D\cdot \hat {\bf T}\cdot \bf d_A  $$ 
with $\hat {\bf T}=\hat {\bf T}^{(0)}+ \hat {\bf T}^{(1)}$, where
$$
\hat {\bf T}^{(0)}(\bm{b})=b^{-3}(3\bf n \otimes \bf n -\hat I)  
\nonumber
$$
is the usual dipole--dipole interaction tensor (${\bf n}={\bf b} / b$) and
\begin{equation} 
\label{eq:T1}
\hat {\bf T}^{(1)}(\bm{b})= \int \! \frac{\mathrm{d}^2q}{(2\pi)^2} \, (-i \bm{e}_q+ \bm{e}_z) \otimes (i \bm{e}_q+\bm{e}_z) q A(q) e^{-2q h+i \bm{q}\cdot \bm{b}} 
\end{equation}
is the part due to the image dipoles. In Equation (\ref{eq:T1}),  $ \bm{e}_q = \bm q /q$, $ \bm{e}_z$ is the unit vector along $z$-axis,
$$
A(q)=\frac{\varepsilon_2 - \varepsilon_1 + f(q)} {\varepsilon_2 + \varepsilon_1+f(q)}\,, \qquad f(q)=i\frac{4 \pi \sigma_g(\omega)}{\omega}q\, ,
\nonumber
$$
and $\sigma_g(\omega)$ is the optical conductivity of graphene. The components of $\hat {\bf T}^{(1)}$ have an oscillatory dependence upon the inter-dot distance~\cite{Huidobro2012}, which arises from the Bessel functions appearing as the result of the angular integration in Equation (\ref{eq:T1}).  
It means that instead of the usual monotonic $b^{-6}$ dependence of the FRET rate one may have an oscillatory behaviour, controllable by the gate voltage through the graphene Fermi energy (which determines $\sigma_g(\omega)$). Beyond the non-monotonic dependence upon the inter-dot distance, the emission/transfer frequency also should affect it in a rather complex way. 
If~this interaction is strong enough, the emitters (QDs in our case) can be coupled to create a collective radiative mode, a phenomenon called superradiance~\cite{Dicke1954}, which has been observed for epitaxial quantum dots~\cite{Scheibner2007}. As the coupling between the emitters can be strongly enhanced or suppressed in the vicinity of graphene, depending on the distance between them, the frequency and the graphene Fermi energy, both superradiance and subradiance regimes can be expected~\cite{Huidobro2012}. So far, we are not aware of an experimental demonstration of such effects. Losses associated with the real part of graphene's optical conductivity can be one possible reason for this.

The influence of a polarisable surface (such as graphene) on the irreversible FRET should be an easier-to-observe effect than the superradiance. It can be seen as a surface plasmons' effect. Huge FRET enhancement factors of the order of $10^6$ (compared to vacuum) have been predicted for two dipole emitters on a graphene sheet, based on numerical electrodynamics calculations~\cite{Agarwal2013}, reaching a maximum when the distance between the dipoles equals twice the graphene plasmon propagation length $L_p$. This prediction is at variance with the results presented in \cite{Huidobro2012} according to which the effect must be considerably smaller and varying on scale of the surface plasmon wavelength, $\lambda _p <<L_p$.    

\section{Concluding Remarks}
\label {conclusion}

To summarize, we would like to emphasize, once again, that the radiative properties and efficiency of light emission in semiconductor nanostructures are determined not only by the radiative transitions, but also by various non-radiative processes which can be even more intense than the radiative transitions themselves. As a result, in these cases, the radiative recombination can be significantly suppressed. On~the other hand, even the non-radiative processes can be used in a constructive way. For instance, the exciton migration is accompanied by a non-radiative energy transfer (FRET) that can be channelled in a desired direction {via} creating a certain ``architecture'' of the nanocrystals in the ensemble, which allows to concentrate and illuminate the energy inside a given area~\cite{Kawazoe,JCTN} or direct it to a certain layer of a funnel-type heterostructure of QD monolayers~\cite{Klar2005}. 
Exciton energy transfer and emission with upconversion because of simultaneous absorption of optical phonons has been suggested for QD-assisted cooling~\cite{Rakovich2009}.  
The multi-exciton effects make it possible to efficiently transform the absorbed photons into the rising number of electron--hole pairs capable of participating in the electric current. Such processes underlie the operating principle of photosynthetic light harvesting systems~\cite{Mirkovic2016} and can be mimicked in solar cells~\cite{Sambur2010}. 

The exciton transport can be strongly enhanced if mediated by polaritons arising under strong coupling of the excitons to light in microcavities~\cite{Coles2014,Schachenmayer2015}.~Combining entities supporting localized excitons, such as QDs, dye molecules or J-aggregates, with microcavities in order to achieve the strong coupling regime is an important area of research envisaging various applications~\cite{Dovzhenko2018}, in particular, in the context of controllable quantum emitters. A plasmonic surface considered in Section \ref {sec_env} can be seen as a near-field microcavity if the lossy channels are not the dominating ones. A detailed discussion of the strong coupling between surface plasmon-polaritons and quantum emitters can be found elsewhere~\cite{Torma2014}.           

\section*{Acknowledgments}
Funding from the Ministry of Science and Higher Education of the Russian Federation (State Assignment No 0729-2020-0058), the European Commission within the project "Graphene-Driven Revolutions in ICT and Beyond" (Ref. No. 696656), from the Portuguese Foundation for Science and Technology (FCT) in the framework of the PTDC/NAN-OPT/29265/2017 "Towards high speed optical devices by exploiting the unique electronic properties of engineered 2D materials" project and the Strategic Funding UID/FIS/04650/2019 is gratefully acknowledged.




\end{document}